\documentclass[onecolumn]{svjour3}          % twocolumn

\usepackage[numbers,comma,sort,compress]{natbib}

\usepackage{listings} % code syntax highlighting

\usepackage{setspace} % for switching between double/single space in document

\usepackage{amsfonts,amssymb,amsbsy,amsmath} %,times,mtpro2} %% times package needed only to load TNR text font}
\usepackage{mathtools}
\usepackage{upgreek}
\usepackage{listings}

\usepackage{graphicx}
\usepackage{xifthen}
\usepackage{bigints}
\usepackage{booktabs}
\usepackage{nicefrac}
\usepackage{color}
\usepackage{url}
\usepackage{multirow}
\usepackage{hyperref} % PDF links
\usepackage[theorems]{tcolorbox}

\usepackage[labelfont=bf]{caption}
\usepackage{subcaption}

\usepackage[a4paper, margin=1in]{geometry}
\usepackage{gensymb} % \degree command

\usepackage{tabularx}
\usepackage{array} % used in tabular env for left alignment of text >{\raggedright}

\usepackage{import} % for nested input files
\usepackage{pifont,latexsym,theorem,rotating,calc} % ,ifthen

\usepackage[]{algorithm2e}
\newcommand{\norm}[1]{\left\lVert#1\right\rVert}

\newcommand{\code}[1]{\texttt{#1}}

\newcommand{\nproc}{{N_p}}
\newcommand{\geodist}{{\mathcal{G}}}
\newcommand{\geoprob}{{\pi_\geodist}}
\newcommand{\accr}{{\alpha}} % mean acceptance rate
\newcommand{\maccr}{{\overline{\accr}}} % mean acceptance rate
\newcommand{\acf}{{\text{ACF}}}
\newcommand{\iac}{{\text{IAC}}}

\newcommand{\Prop}{{Y}}
\newcommand{\prop}{{y}}
\newcommand{\Pam}{{Q}} % proposal during the AM
\newcommand{\pam}{{q}} % proposal during the AM
\newcommand{\pdr}{{g}} % proposal during delayed rejection
\newcommand{\adr}{{\beta}} % acceptance probability during delayed rejection
\newcommand{\pstate}[1][] {{ y^{ \ifthenelse{\isempty{#1}}{}{{#1}} } }}

\newcommand{\dataset}{{\mathcal{\dvar}}}
\newcommand{\datasubset}[1][]{{ \dataset_{ \ifthenelse{\isempty{#1}}{\data}{{\data_{#1}}} } }}

\newcommand{\truthind}[1][]{{ \truth^{ind}_{ \ifthenelse{\isempty{#1}}{}{{#1}} } }}
\newcommand{\truthdep}[1][]{{ \truth^{dep}_{ \ifthenelse{\isempty{#1}}{}{{#1}} } }}
\newcommand{\dataind}[1][] {{  \data^{ind}_{ \ifthenelse{\isempty{#1}}{}{{#1}} } }}
\newcommand{\datadep}[1][] {{  \data^{dep}_{ \ifthenelse{\isempty{#1}}{}{{#1}} } }}

\catcode`_=\active
\newcommand_[1]{\ensuremath{\sb{\mathrm{#1}}}}
\newcommand{\newpar}{{}}

\newcommand{\iid}{{i.i.d.}}
\newcommand{\pdf}[1]{{\pi(#1)}}

\newcommand{\up}{\operatorname}
\newcommand{\diff}{{\up{d}}}

\newcommand{\bs}{\boldsymbol}

\newcommand{\numChar}{{n}} % character representing Number
\newcommand{\paraChar}{{p}}

\newcommand{\episChar}{{n}}

\newcommand{\npe}[1][]{{ \numChar_{ \ifthenelse{\isempty{#1}}{\paraChar\episChar}{{\paraChar\episChar,#1}} } }} % number of epistemic parameters in the model
\newcommand{\epis}{{nois}}

\newcommand{\scen}{{\mathcal{S}}}

\def\gtrsim{\mathrel{\hbox{\rlap{\hbox{\lower4pt\hbox{$\sim$}}}\hbox{$>$}}}}
\def\lessim{\mathrel{\hbox{\rlap{\hbox{\lower4pt\hbox{$\sim$}}}\hbox{$<$}}}}

\newcommand{\rmz}{{\rm z}}

\newcommand{\emodel}[1][]{{ M_{ \ifthenelse{\isempty{#1}}{\epis}{{\epis,#1}} } }}
\newcommand{\emodelz}[1][]{{ M_{ \rmz\ifthenelse{\isempty{#1}}{\epis}{{\epis,#1}} } }}
\newcommand{\emodellgrb}[1][]{{ M^\lgrb_{ \ifthenelse{\isempty{#1}}{\epis}{{\epis,#1}} } }}

\newcommand{\lgrb}{{\rm g}}

\newcommand{\param}{{\bs{\theta}}}
\newcommand{\eparam}[1][]{{ \param_{ \ifthenelse{\isempty{#1}}{\epis}{{\epis,#1}} } }}
\newcommand{\eparamz}[1][]{{ \bs\param^\rmz_{ \ifthenelse{\isempty{#1}}{\epis^\rmz}{{\epis,#1}} } }}
\newcommand{\eparamlgrb}[1][]{{ \bs\param^\lgrb_{ \ifthenelse{\isempty{#1}}{\epis^\lgrb}{{\epis,#1}} } }}
\newcommand{\truth}{{\bs{R}}}
\newcommand{\possible}{{*}}
\newcommand{\truthset}{{\mathcal{R}}}

\newcommand{\truthsubset}[1][]{{ \truthset_{ \ifthenelse{\isempty{#1}}{\truth}{{\truth_{#1}}} } }}
\newcommand{\ptruthsubset}[1][]{{ \truthset_{ \ifthenelse{\isempty{#1}}{\truth}{{\truth_{#1}}} }^\possible }}

\newcommand{\dset}{{\mathcal{X}}}
\newcommand{\temp}{{\mathcal{T}}}

\DeclareMathOperator*{\argmax}{\operatorname*{arg\,max}}

\newcommand{\xx}[1][]{{ \ifthenelse{\isempty{#1}}{\textcolor{red}{XXX}}{\textcolor{red}{~(XXX {#1} XXX)~}} }}

\defcitealias{fermi2018gamma}{F18}

\smartqed  % flush right qed marks, e.g. at end of proof

\theoremstyle{remark}

\usepackage{array}

\makeatletter
\newcommand{\thickhline}{%
    \noalign {\ifnum 0=`}\fi \hrule height 2pt
    \futurelet \reserved@a \@xhline
}
\newcolumntype{"}{@{\hskip\tabcolsep\vrule width 2pt\hskip\tabcolsep}}
\makeatother

\graphicspath{
{./fig/intro/}
{./fig/probability/}
{./fig/challenges/}
{./fig/optimization/}
{./fig/sampling/}
{./fig/integration/}
} %Where the figures folder is located

\begin{document}

%%%%%%%%%%%%%%%%%%%%%%%%%%%%%%%%%%%%%%%%%%%%%%%%%%%%%%%%%%%%%%%%%%%%

\setlength{\parindent}{0pt}
\newlength{\stretchlen}\setlength{\stretchlen}{1em}
\def\splitterm{\_}
\newcommand{\eqd}[1]{\leavevmode\realstretch#1\_}
\def\realstretch#1{%
    \def\temp{#1}%
    \ifx\temp\splitterm
    \else
    \hbox to \stretchlen{\hss#1\hss}\expandafter\realstretch
\fi}
%\eqd{abcd}\par

%%%%%%%%%%%%%%%%%%%%%%%%%%%%%%%%%%%%%%%%%%%%%%%%%%%%%%%%%%%%%%%%%%%%

\onehalfspacing
\large
\title{\bf \Large {ParaDRAM: A Cross-Language Toolbox for Parallel High-Performance Delayed-Rejection Adaptive Metropolis Markov Chain Monte Carlo Simulations}}
%\subtitle{Computational Techniques for Scientific Inference}

\titlerunning{\large Parallel Delayed-Rejection Adaptive MCMC}        % if too long for running head

\author{\large Amir Shahmoradi$^{1}$ \and Fatemeh Bagheri$^{1}$}
\authorrunning{\large Amir Shahmoradi and Fatemeh Bagheri} % if too long for running head

\institute{\normalsize $^1$ Department of Physics, \\
           $^{~}$ Data Science Program, College of Science \at
           $^{~}$ The University of Texas, Arlington, TX 76010 \\
           $^{~}$ \email{a.shahmoradi@uta.edu} \\
           $^{~}$ \email{fatemeh.bagheri@uta.edu} \\
           %Tel.: +123-45-678910 \\
           %Fax: +123-45-678910 \\
           %\emph{Present address:} of F. Author  %  if needed
           % \and
           % $^\dagger$ Peter O'Donnell, Jr. Fellow
}

\date{Received: date / Accepted: date}
% The correct dates will be entered by the editor

\maketitle

\begin{abstract}
We present ParaDRAM, a high-performance {\bf Para}llel {\bf D}elayed-{\bf R}ejection {\bf A}daptive {\bf M}etropolis Markov Chain Monte Carlo software for optimization, sampling, and integration of mathematical objective functions encountered in scientific inference. ParaDRAM is currently accessible from several popular programming languages including C/C++, Fortran, MATLAB, Python and is part of the ParaMonte open-source project with the following principal design goals: 1. {\bf full automation} of Monte Carlo simulations, 2. {\bf interoperability} of the core library with as many programming languages as possible, thus, providing a unified Application Programming Interface and Monte Carlo simulation environment across all programming languages, 3. {\bf high-performance} 4. {\bf parallelizability} and scalability of simulations from personal laptops to supercomputers, 5. virtually {\bf zero-dependence on external libraries}, 6. {\bf fully-deterministic reproducibility} of simulations, 7. {\bf automatic comprehensive reporting} and post-processing of the simulation results. We present and discuss several novel techniques implemented in ParaDRAM to automatically and dynamically ensure the good-mixing and the diminishing-adaptation of the resulting pseudo-Markov chains from ParaDRAM. We also discuss the implementation of an efficient data storage method used in ParaDRAM that reduces the average memory and storage requirements of the algorithm by, a factor of 4 for simple simulation problems to, an order of magnitude and more for sampling complex high-dimensional mathematical objective functions. Finally, we discuss how the design goals of ParaDRAM can help users readily and efficiently solve a variety of machine learning and scientific inference problems on a wide range of computing platforms.
\end{abstract}

%\clearpage
%\onehalfspacing
\tableofcontents
\section{Introduction}
\label{sec:intro}

\begin{figure}[t!]
        \centering
        \begin{subfigure}[t]{0.49\textwidth}
            \centering
            \includegraphics[width=\textwidth]{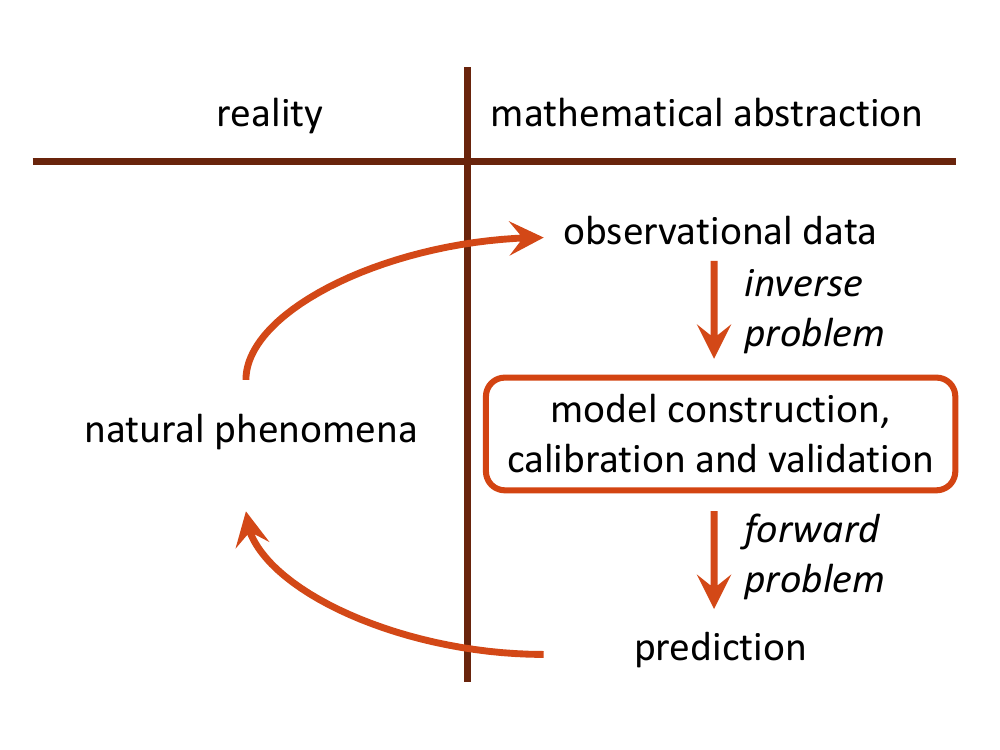}
            \caption{The scientific methodology.} \label{fig:scientificMethod}
        \end{subfigure}
        \begin{subfigure}[t]{0.49\textwidth}
            \centering
            \includegraphics[width=\textwidth]{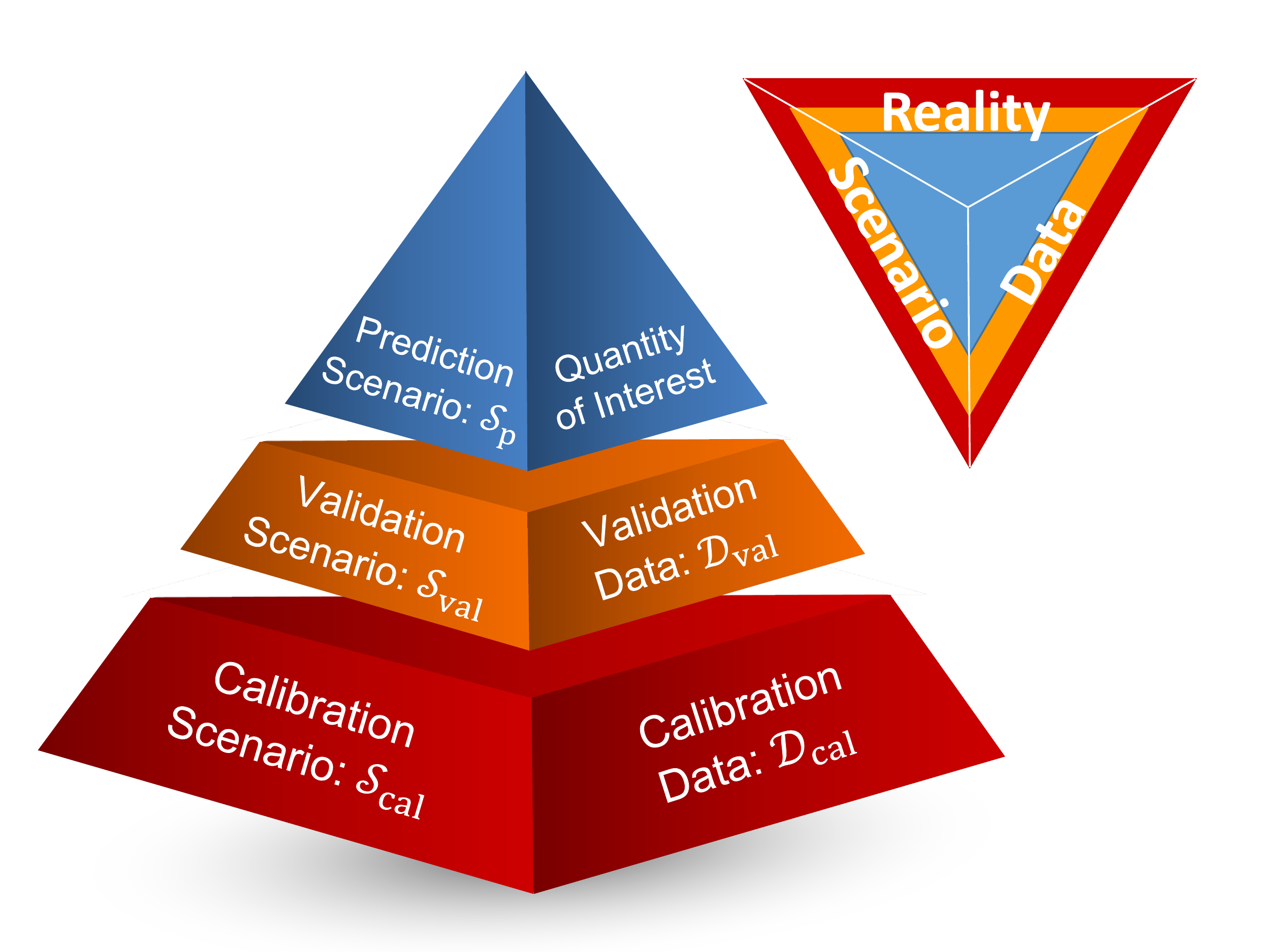}
            \caption{The prediction pyramid.} \label{fig:pyramid}
        \end{subfigure}
        \caption{{\bf(a)} {\bf An illustration of the fundamental steps in predictive science}, which includes data collection, hypothesis formulation, as well as the construction of a mathematical model and objective function, which is subsequently optimized to constrain the parameters of the model in a process known as {\it inversion} or {\it inverse problem}. Once validated, the model can be used to make predictions about the quantity of interest ({\it forward problem}). {\bf(b)} {\bf The prediction pyramid}, depicting the three hierarchical levels of predictive inference from bottom to top: Calibration, Validation, and Prediction of the Quantity of Interest (QoI). The rear face of the tetrahedron represents reality (truth), $\truthset$, about the set of observed phenomena, which is never known to the observer. The front-right face of the tetrahedron represents the observational data, $\dset$, which results from the convolution of the truth/reality, $\truthset$, with various forms of measurement uncertainty. The front-left face represents the scenarios, $\scen$, under which data is collected, as well as the set of models that are hypothesized to describe the unknown truth, $\truthset$ \cite{oden2004predictive, oden2010computer, oden2013selection, oden2017}. (\scriptsize Adapted from \citep{shahmoradi2017multilevel, 2017arXiv171110599S, oden2018adaptive})}
    \end{figure}

    At the very foundation of predictive science lies the scientific method which involves multiple steps of making observations and developing testable hypotheses and theories of natural phenomena. Once a scientific theory is developed, it can be cast into a mathematical model which then serves as a proxy-seek of the truth. Then, the free parameters of the model, if any, has to be constrained, or {\it calibrated}, using the {\it calibration data} in a process known as the {\it model calibration} or the {\it inverse-problem}. The validity of the model -- and thereby, the scientific theory behind the model -- are subsequently tested against a {\it validation dataset} that has been collected independently of the calibration data. Once the model is calibrated and validated, it can be used to predict the quantity of interest (QoI) of the problem in a process known as the {\it forward-problem} \citep{oden2018adaptive}. This entire procedure is schematically illustrated in Figure \ref{fig:scientificMethod}. The hierarchy of model calibration, validation, and prediction has become known as the {\it prediction pyramid}, as depicted in Figure \ref{fig:pyramid} \citep{oden2004predictive, oden2010computer, oden2013selection, oden2017}.

    A major task in the calibration step of every scientific prediction problem is to find the best solution -- among the potentially infinite set of all possible solutions -- to a mathematically-defined problem, where the mathematical model serves as an abstraction of the physical reality (Figure \ref{fig:scientificMethod}). Specifically, finding the best solution (i.e., the best-fit model parameters) requires,

    \begin{enumerate}
        \item the construction of one (or more) mathematical objective function(s) that quantifies the goodness of each set of possible parameters for the model, and then,
        \item optimizing the objective function(s) (for {\bf parameter tuning}), and$/$or,
        \item sampling the objective function(s) (for {\bf uncertainty quantification}), and$/$or,
        \item integrating the objective function(s) (for {\bf model selection}).
    \end{enumerate}

    In this process, optimization is primarily performed to obtain the best-fit parameters of the model given the calibration dataset. The history of mathematical optimization dates back to the emergence of modern science during the Renaissance, perhaps starting with a paper by Pierre de Fermat in 1640s on finding the local extrema of differentiable functions \citep[][]{mahoney1994mathematical, du2013mathematical}. A second revolution in the field of optimization occurred with the (re-)discovery \citep{floudas2001encyclopedia} of linear programming by Dantzig in 1947 \citep{dantzig1998linear}, followed by the first developments in {\it nonlinear programming} \citep{kuhn1951nonlinear}, {\it stochastic programming} \citep{dantzig1955linear, beale1955minimizing}, and the revival of interest in {\it network flows}, {\it combinatorial optimization} \citep{ford1955simple} and {\it integer programming} \cite{gomory1963algorithm} long after the original works of Fermat in the $17^{th}$ century.\newpar

    Independently of the rapid developments in the field of mathematical programming, a new branch of science began to sprout in the late 1940s at Los Alamos National Laboratory (LANL), resulting, most notably, from the early works of Enrico Fermi \citep{segre1955fermi, metropolis1987beginning}, Stanislaw Ulam, John Von Neumann, along with Edward Teller, Marshall Rosenbluth, and Nicholas Metropolis. In a series of articles \citep[e.g.,][]{metropolis1949monte, von195113, metropolis1953equation} they form the foundations of what becomes known in the following decades as {\it stochastic simulation} or {\it Monte Carlo methods}\footnote{The technique's name `Monte Carlo' was a suggestion made by Metropolis not so unrelated to Stan Ulam's uncle who used to borrow money from relatives because he ``just had to go to Monte Carlo" for gambling \citep{metropolis1987beginning}.}. In their works, the authors propose several seminal methods for sampling strictly non-negative-valued generic mathematical functions in arbitrary dimensions, but mostly in the context of problems encountered in the field of Statistical Physics.
    \newpar

    The proposed methodology of \citet{metropolis1953equation} for sampling mathematical density functions was perhaps not fully appreciated by the scientific community until \citet{hastings1970monte} presented a more generic formulation of the proposed sampling approach of \cite{metropolis1953equation}, now known as the {\it Metropolis-Hastings Markov Chain Monte Carlo} ({\bf MH-MCMC}) method. These two monumental articles, along with rapid technological breakthroughs in the world of computers have now enabled researchers to achieve all three aforementioned fundamental goals of the predictive science ({\it parameter-tuning}, {\it uncertainty-quantification}, and {\it model-selection}) in their research.
    \newpar

    Optimization and Monte Carlo techniques have played a fundamental role in the emergence of the third pillar of science, {\it computational modeling} \citep{oden2010computer, shahmoradi2017multilevel}, in the 1960s alongside the two original pillars of science: {\it observation} and {\it theory}. In particular, the MCMC techniques have become indispensable practical tools across all fields of science, from Astrophysics and Climate Physics \citep[e.g., ][]{shahmoradi2013multivariate,shahmoradi2015short, 2020arXiv200601157O, 2019arXiv190306989S, li2019uncertainty, curcic2019parallel} to Bioinformatics and Biomedical Sciences \citep[e.g., ][]{shahmoradi2014predicting, lima2017ices, lima2017selection, jha2020bayesian} or Engineering fields \citep[e.g.,][]{oden2017predictive, taghizadeh2020bayesian}.
    \newpar

    Despite their popularity, the MCMC methods, in their original form as laid out by \citet{hastings1970monte}, have a significant drawback: The methods often require hand-tuning of several parameters within the sampling algorithms to ensure fast convergence of the resulting Markov chain to the target density for the particular problem at hand. Significant research has been done over the past decades, in particular during 1990' and 2000' to bring automation to the problem of tuning the free parameters of the MCMC methods. Among the most successful attempts is the algorithm of \citet{haario2006dram}, known as the {\bf D}elayed-{\bf R}ejection {\bf A}daptive {\bf M}etropolis MCMC ({\bf DRAM}).
    \newpar

    Several packages already provide implementations of variants of the proposed DRAM algorithm in \citet{haario2006dram}. Peer-reviewed open-source examples include \code{FME} \citep{soetaert2010inverse} in R, \code{PyMC} \citep{patil2010pymc} in Python, \code{mcmcstat} \citep{haario2006dram} in MATLAB, \code{mcmcf90} \citep{haario2006dram} in Fortran, and QUESO \citep{queso, mcdougall2015parallel} in C/C++ programming languages. However, despite implementing the same algorithm (DRAM), these packages dramatically differ in their implementation approach, Application Programming Interface (API), computational efficiency, parallelization, and accessibility from a specific programming environment.
    \newpar

    The ParaDRAM algorithm presented in this manuscript attempts to address the aforementioned heterogeneities and shortcomings in the existing implementations of the DRAM algorithm by providing a unified Application Programming Interface and environment for MCMC simulations accessible from multiple programming languages, including C/C++, Fortran, MATLAB, Python, R, with ongoing efforts to support other popular contemporary programming languages.

    The ParaDRAM algorithm is part of the open-source Monte Carlo simulation library with a codebase currently comprised of approximately $130,000$ lines of code in mix of programming languages, including C, Fortran, MATLAB, Python, R, as well as Bash, Batch, and CMake scripting languages and build environments. The ParaDRAM package has been designed while bearing the following design philosophy and goals in mind,
    \begin{enumerate}
        \item
            {\bf Full automation} of all Monte Carlo simulations to ensure the highest level of user-friendliness of the library and minimal time investment requirements for building, running, and post-processing of MCMC simulations.
        \item
            {\bf Interoperability} of the core library with as many programming languages as currently possible, including C/C++, Fortran, MATLAB, Python, R, with ongoing efforts to support other popular programming languages.
        \item
            {\bf High-Performance} meticulously-low-level implementation of the library to ensure the fastest-possible Monte Carlo simulations.
        \item
            {\bf Parallelizability} of all simulations via two-sided and one-sided MPI/Coarray communications while requiring zero-parallel-coding efforts by the user.
        \item
            {\bf Zero-dependence} on external libraries to ensure hassle-free ParaDRAM simulation builds and runs.
        \item
            {\bf Fully-deterministic reproducibility} and automatically-enabled restart functionality for all simulations up to 16 digits of precision if requested by the user.
        \item
            {\bf Comprehensive-reporting and post-processing} of each simulation and its results, as well as their automatic compact storage in external files to ensure the simulation results will be comprehensible and reproducible at any time in the distant future.
    \end{enumerate}

    As implied by its name, a particular focus in the design of the ParaDRAM algorithm is to ensure seamlessly-scalable parallelizations of Monte Carlo simulations, form personal laptops to supercomputers, while requiring absolutely no parallel-coding effort by the user. In the following sections, we will describe the design, implementation, and algorithmic details and capabilities of ParaDRAM. Toward this, we will devote \S\ref{sec:mhmcmc}, \S\ref{sec:dram}, and \S\ref{sec:paradram} on the mathematical explanation of the algorithm, including our proposed approach to dynamic monitoring of the diminishing-adaption condition of the DRAM algorithm (\S\ref{sec:paradram:diminishing}), as well as the parallelization paradigms used in the ParaDRAM algorithm (\S\ref{sec:paradram:parallelism}). Then, we describe the Application Programming interface of ParaDRAM in \S\ref{sec:api}, including the implementation details of some of the unique features of the algorithm that enhances its computational and memory usage efficiency. Finally, we discuss the practical performance of the ParaDRAM algorithm in \S\ref{sec:results} and the road ahead for extending this package in \S\ref{sec:discussion}.

\section{The Metropolis-Hastings MCMC algorithm}
\label{sec:mhmcmc}

    The original proposed approach to sampling a mathematical objective density function, $f(x)$, by \citep{metropolis} and \citep{hastings1970monte} is based on the brilliant observation that a special type of discrete-time stochastic processes, known as Markov process or Markov chain, can have stationary distributions under some conditions. Suppose a stochastic sequence of random vectors,

    \begin{equation}
        \label{eq:randomSeq}
        \{X_i:~i = 1, \dots, +\infty\} ~,
    \end{equation}

    \noindent is sampled from the $d$-dimensional domain of the objective function representing the state-space $\dset$, such that it possesses a unique feature known as the Markov property. For notational simplicity, we assume that $\dset$ is a finite set. The Markov property of this chain requires that, given the present state, the future random state of the sequence must be independent of the past states,

    \begin{equation}
        \label{eq:markovProperty}
        \pdf{
            X_{i+1} = x_{i+1} |
            X_{i} = x_{i},
            \dots,
            X_{1} = x_{1},
        } =
        \pdf{
            X_{i+1} = x_{i+1} |
            X_{i} = x_{i}
        } ~.
    \end{equation}

    Here $\pdf{\cdot}$ denotes the probability. The entire Markov process is characterized by an initial state as well as a square {\it transition matrix}, $P$, whose elements, $P_{ij}$, describe the probabilities of transitions from each state $i$ to every other possible state $j$ in $\dset$,

    \begin{equation}
        \label{eq:transitionMatrixElement}
        P_{ij} =
        \pdf{
            X_{n+1} = j |
            X_{n} = i
        } ~~ \forall ~ n \in \{1,\dots,+\infty\} ~.
    \end{equation}

    If there is a unique transition matrix for the Markov process, then it is a {\it time-homogenous} Markov chain. Furthermore, if the Markov process is ergodic (i.e., aperiodic and $\Phi$-irreducible, that is, capable of visiting every state in a finite time \citep{roberts2007coupling}) and, there is a distribution $f(x)$ such that every possible transition $x \rightarrow \prop$ in the process follows the principle of {\it detailed balance},

    \begin{equation}
        \label{eq:detailedBalance}
        f(x) \pdf{ \prop | x } = f(\prop) \pdf{ x | \prop } ~,
    \end{equation}

    \noindent then, the process can be shown to have a unique {\it stationary distribution}, $f(x)$, to which it asymptotically approaches. The challenge, however, is to find a transition matrix, $\pdf{ \prop | x }$, that obeys the above conditions with respect to the target objective density function of interest, $f(x)$. The revolutionary insight due to \citep{metropolis} and \citep{hastings1970monte} is that, one can define such generic transition matrix with respect to the desired $f(x)$, fulfilling the above conditions, if the transition matrix is split into two separate terms:

    \begin{enumerate}
        \item
            a proposal step, during which one proposes a new state $\prop$ distributed according to $\pam(\prop|x)$, given the current state $x$,
        \item
            followed by the acceptance or the rejection of the proposed step according to,
            \begin{equation}
                \label{eq:acceptanceRate}
                \accr(x,\prop) = \min\bigg( 1 , \frac{f(\prop)\pam(x|\prop)}{f(x)\pam(\prop|x)} \bigg) ~,
            \end{equation}
    \end{enumerate}

    \noindent such that the transition probability can be written as,

    \begin{equation}
        \label{eq:transitionPorbabilityHastings}
        \pi(\prop|x) = \pam(\prop|x) \accr(x,\prop) + \delta_{x}(\prop) \sum_{z\in \mathcal{X}} \big(1-\accr(x,z)\big) ~ \pam(z|x) ~,
    \end{equation}

    \noindent where $\delta_x(\prop)$ is an indicator function, a discrete equivalent of the Dirac measure, such that $\delta_x(\prop=x)=1$. Defining the transition probability according to \eqref{eq:transitionPorbabilityHastings} is sufficient, though not necessary \citep[e.g.,][]{barker1965monte}, to guarantee the asymptotic convergence of the distribution of the resulting Markov chain to the target objective density function, $f(x)$ \citep{chib1995understanding, tierney1998note}.
    \newpar

\begin{figure}[t!]
        \centering
        \begin{subfigure}[t]{0.32\textwidth}
            \centering
            \includegraphics[width=\textwidth]{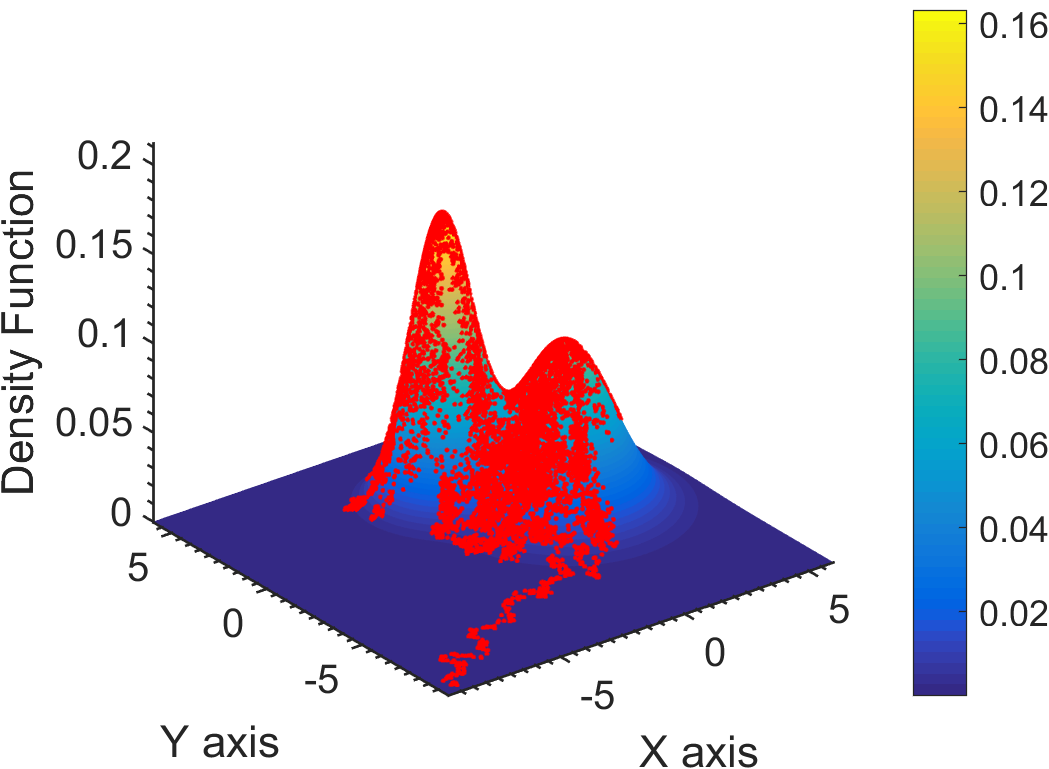}
            \caption{Small-step-MCMC sampling.} \label{fig:smallStepSampling}
        \end{subfigure}
        \hfill
        \begin{subfigure}[t]{0.32\textwidth}
            \centering
            \includegraphics[width=\textwidth]{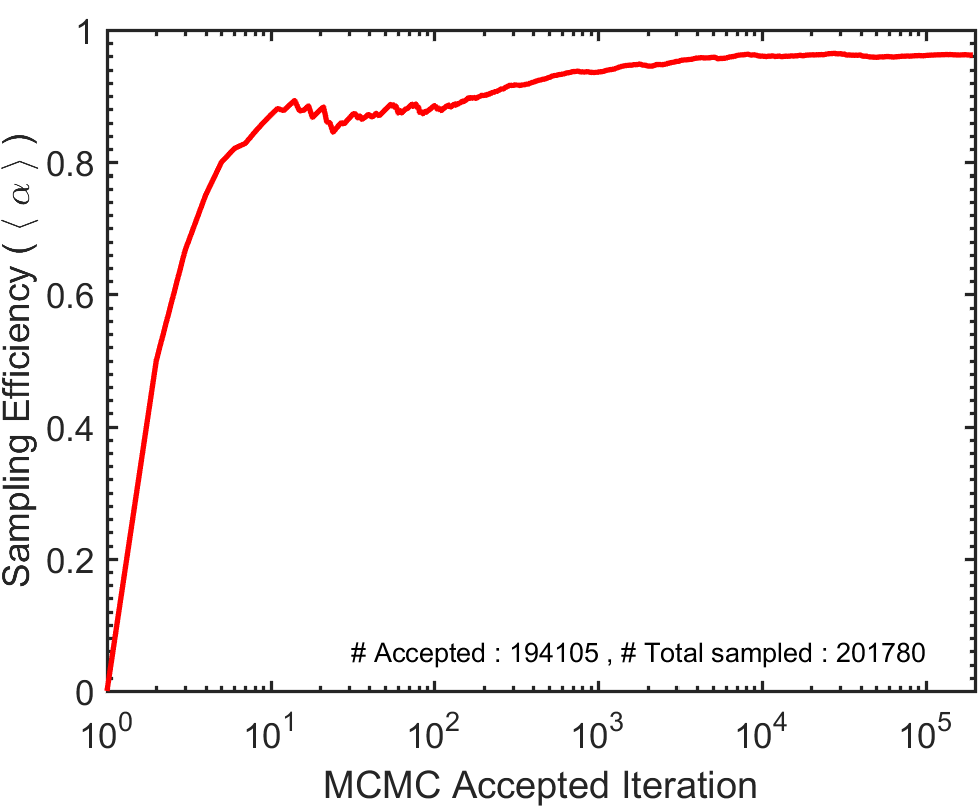}
            \caption{Small-step-MCMC efficiency.} \label{fig:smallStepEfficiency}
        \end{subfigure}
        \hfill
        \begin{subfigure}[t]{0.32\textwidth}
            \centering
            \includegraphics[width=\textwidth]{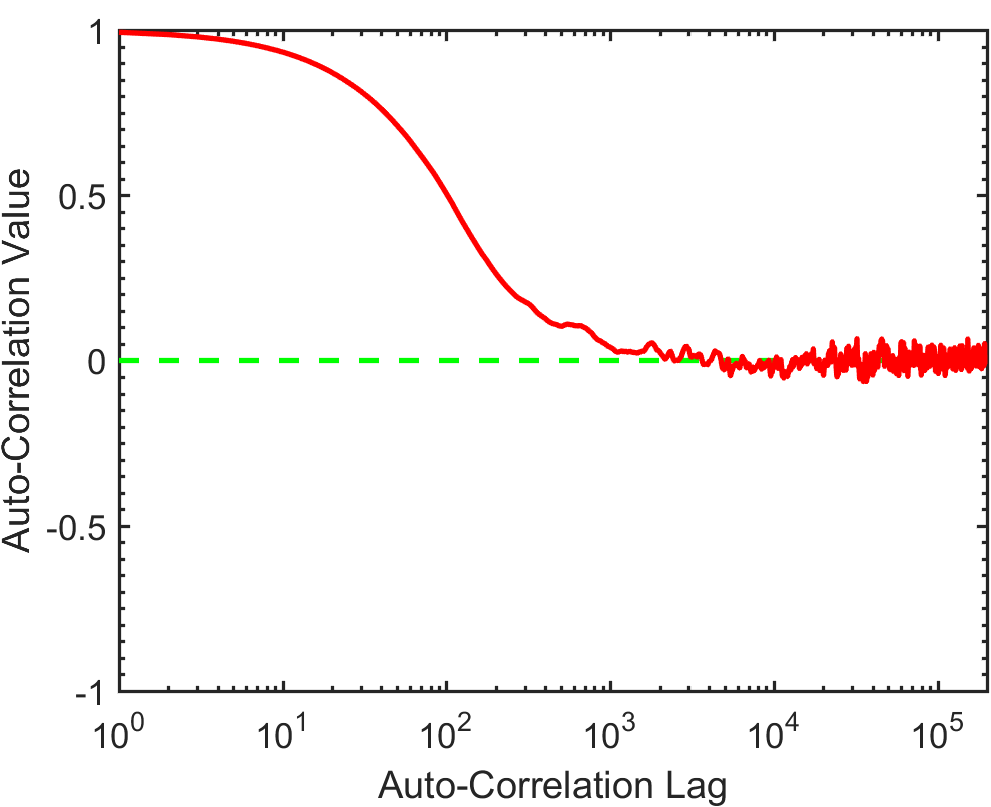}
            \caption{Small-step-MCMC ACF.} \label{fig:smallStepAutoCorr}
        \end{subfigure}
        \hfill
        \begin{subfigure}[t]{0.32\textwidth}
            \centering
            \includegraphics[width=\textwidth]{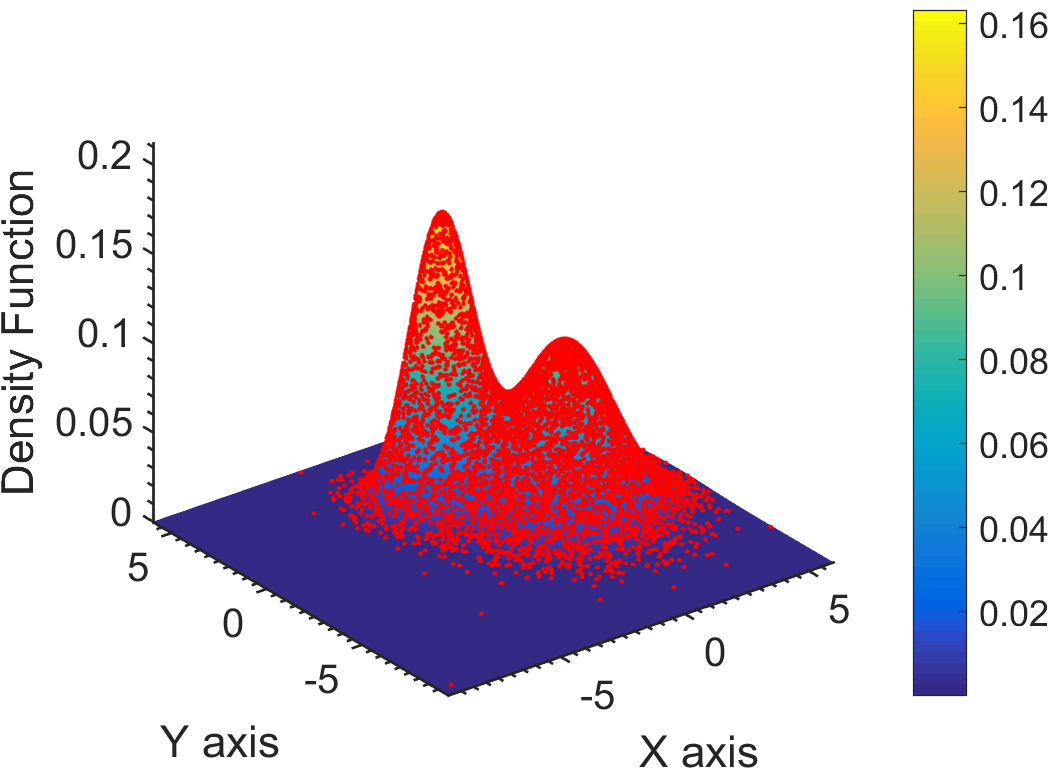}
            \caption{Large-step-MCMC sampling.} \label{fig:largeStepSampling}
        \end{subfigure}
        \hfill
        \begin{subfigure}[t]{0.32\textwidth}
            \centering
            \includegraphics[width=\textwidth]{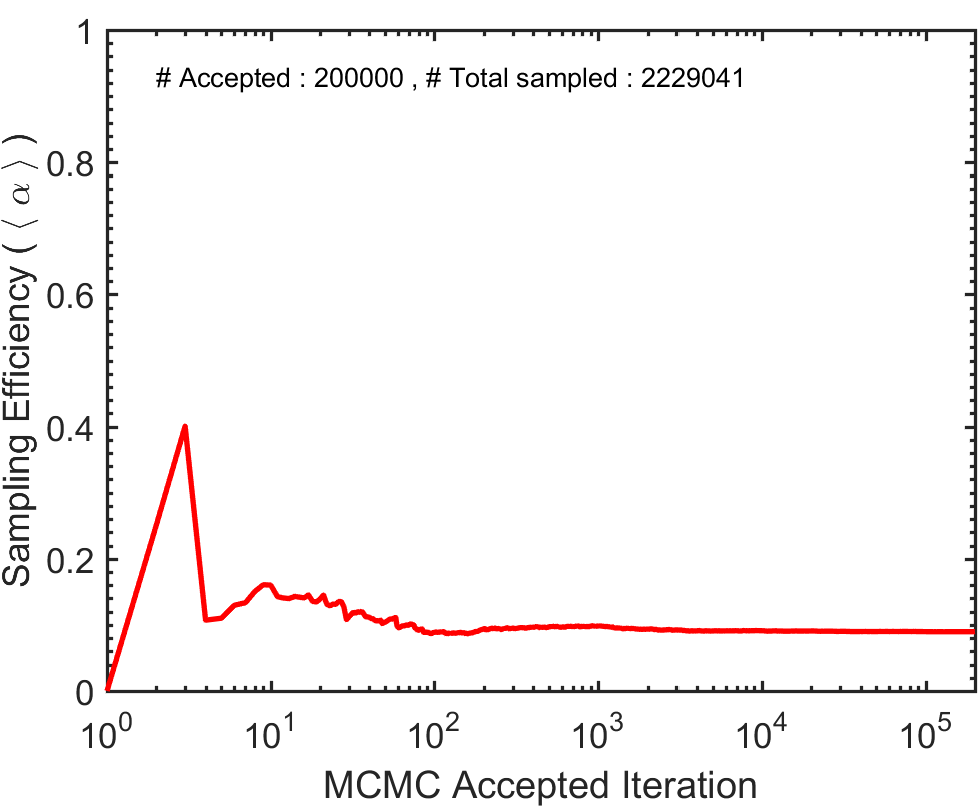}
            \caption{Large-step-MCMC efficiency.} \label{fig:largeStepEfficiency}
        \end{subfigure}
        \hfill
        \begin{subfigure}[t]{0.32\textwidth}
            \centering
            \includegraphics[width=\textwidth]{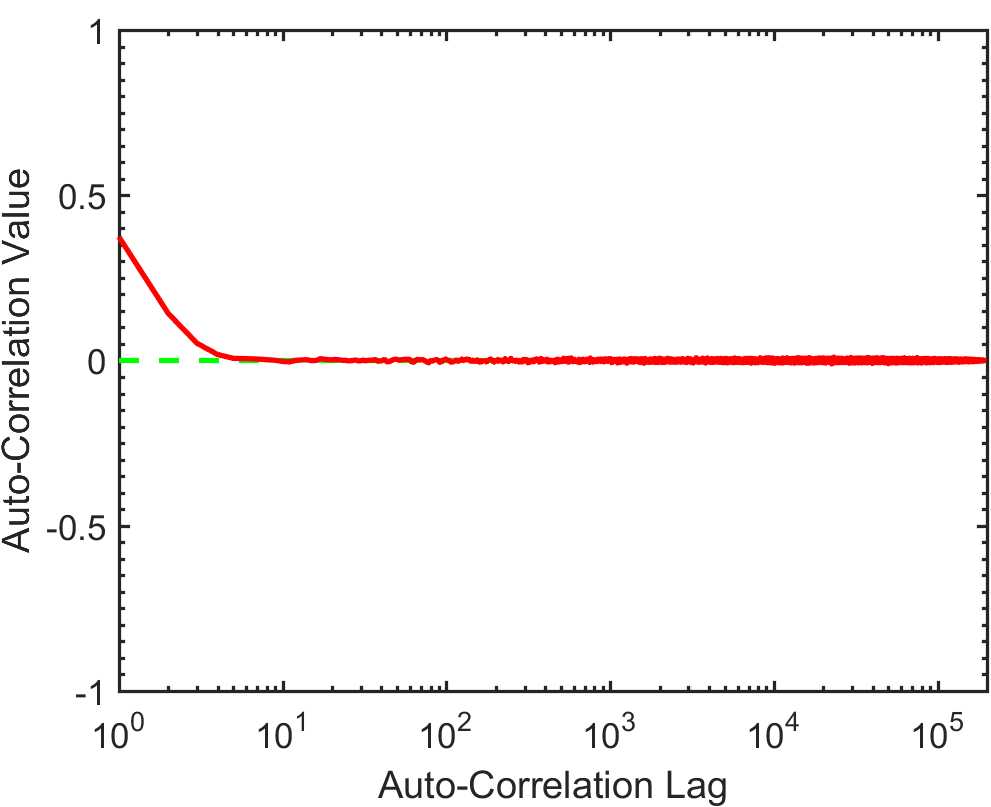}
            \caption{Large-step-MCMC ACF.} \label{fig:largeStepAutoCorr}
        \end{subfigure}
        \caption{{\bf An illustration of the importance of an appropriate choice of step size and proposal distribution shape in MCMC simulations.} The plots {\bf(a)}, {\bf(b)}, and {\bf(c)} display respectively the MCMC sample, the evolution of the efficiency of the MCMC simulation as defined by \eqref{eq:averageAcceptanceRate}, and the autocorrelation function (ACF) of the chain of {\it uniquely-sampled} states via a proposal distribution whose scale is very small compared to the scale of the target density function. The plots {\bf(d)}, {\bf(e)}, and {\bf(f)} represent the same quantities respectively as in the top plots, however, for very large-step-size proposed moves.\label{fig:stepSizeEffects}}
    \end{figure}

    In practice, however, convergence to the target density can be extremely slow if the proposal distribution, $\pam(\prop|x)$, is too different from $f(x)$ \citep{rosenthal2011optimal}, as illustrated in Figure \ref{fig:stepSizeEffects}. Theoretical results \citep[e.g.,][]{gelman1996efficient} indicate that an average acceptance rate of,

    \begin{eqnarray}
        \label{eq:averageAcceptanceRate}
        \langle\accr\rangle
        = \frac{\text{the number of accepted MCMC proposed moves}}{\text{the total number of accepted and rejected proposed moves}}
        = 0.234 ~,
    \end{eqnarray}

    \noindent yields the fastest convergence rate in the case of an infinite-dimensional standard MultiVariate Normal (MVN) target density function, although numerical experiments indicate the validity of the results to hold for as low as 5 dimensions. In the absence of a generic universal rule for optimal sampling, a common practice has been to adapt the shape of $\pam(\prop|x)$ to the shape of $f(x)$ repeatedly and manually \citep{andrieu2006ergodicity} such that the autocorrelation of the resulting MCMC chain is minimized, but this approach is cumbersome and difficult for large-scale simulation problems.
    \newpar

%%%%%%%%%%%%%%%%%%%%%%%%%%%%%%%%%%%%%%%%%%%%%%%%%%%%%%%%%%%%%%%%%%%%%%%%%%%%%%%%%%%%%%%%%%%%%%%%%%%%%%%%%%%%%%%%%%%%%%%%%%%%%%%%%%%

\section{The DRAM algorithm}
\label{sec:dram}

    A second major insight, due to \citet{haario2001adaptive}, is that one can progressively and continuously adapt the shape of the proposal distribution based on the currently-sampled points in the entire history of the chain. Although the resulting chain is not Markovian because of the explicit dependence of every new sampled point on all of the past visited points, \citet{haario2001adaptive} prove the asymptotic convergence of the resulting chain to the target density, $f(x)$. \citet{roberts2007coupling} also provide more generic conditions under which the ergodicity of the resulting adaptive chain is maintained. The ergodicity property is one of the two pillars upon which the Metropolis-Hastings Markov Chain Monte Carlo is built.
    \newpar

    Compared with the traditional MH-MCMC algorithm, the $i$th-stage acceptance probability of the Adaptive algorithm is modified to the following,

    \begin{equation}
        \label{eq:acceptanceRateAM}
        \accr(x_i,\prop) = \min\bigg( 1 , \frac{f(\prop)q_i(x_i|\prop,x_{i-1},\ldots,x_1)}{f(x)q_i(\prop|x_i,x_{i-1},\ldots,x_1)} \bigg) ~,
    \end{equation}

    A major requirement for the validity of the Adaptive algorithm results is the condition of {\it diminishing adaptation}, requiring that the difference between the adjacent adapted Markov chain kernels approaches to zero in probability as the sample size grows to infinity. A practical implementation of monitoring this condition in the ParaDRAM algorithm is discussed later in \S\ref{sec:paradram:diminishing}.
    \newpar

    \citet{haario2006dram} further combine the adaptive Metropolis algorithm of \citet{haario2001adaptive} with the Delayed-Rejection (DR) algorithm of \citet{green2001delayed} to introduce the Delayed-Rejection Adaptive Metropolis (DRAM) algorithm. In brief, the DR algorithm modifies the standard MH-MCMC to improve the efficiency of the resulting estimators, with the basic idea being that, upon rejecting a proposed state according to \eqref{eq:acceptanceRate}, instead of advancing the chain and retaining the same position as it is done in the MH algorithm, a second-stage move is proposed given the knowledge of the newly-rejected proposed state, perhaps using an entirely different proposal.
    \newpar

    The acceptance probability of the proposed state at each level of the DR process can computed such that the reversibility of the Markov chain relative to the target density $f(x)$ is preserved. Denote, respectively, the proposal Probability Density Function (PDF), the proposed state, and the corresponding acceptance probability at the zeroth stage of the delayed-rejection of the $i$th MCMC step of the DRAM algorithm by $\pdr_0(\prop_0|x) = \pam_i(\prop_0|x)$, $\prop_0$, and $\adr_0(x,\prop_0)$. This zeroth stage is the regular Adaptive Metropolis (AM) algorithm step during which no delayed-rejection occurs. If $\prop_0$ is rejected, another proposal is made at the first level of DR via a potentially-new proposal with the PDF $\pdr_1(\cdot)$. The corresponding acceptance probability is,

    \begin{equation}
        \label{eq:acceptanceRate1}
        \adr_1(x,\prop_0,\prop_1) = \min
        \bigg(
            1 ,
            \frac{f(\prop_1)}{f(x)} \times
            \frac{ \pdr_0(\prop_0|\prop_1) \pdr_1(x|\prop_0,\prop_1) }{ \pdr_0(\prop_0|x) \pdr_1(\prop_1|\prop_0,x) } \times
            \frac{ \big( 1 - \adr_0(\prop_1,\prop_0) \big) }{ \big( 1 - \adr_0(x,\prop_0) \big) }
        \bigg) ~,
    \end{equation}

    Here, $\prop_0$ represents the original AM-proposed state, generated by the AM proposal kernel with the PDF $\pam(\cdot)$. Once $\prop_0$ is rejected, the first DR-proposed state $\prop_1$ is generated according to $\pdr_1(\cdot)$, whose acceptance is computed according to \eqref{eq:acceptanceRate1}.
    \newpar

    The process of delaying the rejection can be continued for as long as desired at every step of the MH algorithm, with the higher-stage delayed-rejection proposals being allowed to depend on the candidates so far proposed and rejected. Since the entire process is designed to preserve the detailed-balance condition of \eqref{eq:detailedBalance}, any acceptance at any stage of the DR process represents a valid MCMC sampling. However, continuing the DR process is feasible only at the cost of increasingly more complex equations for the acceptance probabilities at higher stages of the DR with lower-values. In general, the acceptance rate for the $j$th stage of the delayed rejection process is,

    \begin{eqnarray}
        \label{eq:acceptanceRatej}
        &&\adr_j(x,\prop_0,\prop_1,\ldots,\prop_j) = \nonumber \\
        && \min \bigg( 1 , \frac{f(\prop_j)}{f(x)} \times \frac
            { \pdr_0(\prop_{j-1}|\prop_{j}) \pdr_1(\prop_{j-2}|\prop_{j-1}, \prop_{j}) \cdots \pdr_j(x|\prop_{1}, \ldots, \prop_{j}) }
            { \pdr_0(\prop_0|x) \pdr_1(\prop_1|\prop_0,x) \cdots  \pdr_i(\prop_1|\prop_i,\ldots,\prop_0,x) } \times \nonumber \\
        && \frac
            { \big( 1 - \adr_0(\prop_j,\prop_{j-1}) \big) \big( 1 - \adr_1(\prop_j,\prop_{j-1},\prop_{j-2}) \big) \cdots \big( 1 - \adr_{j-1}(\prop_j,\prop_{j-1},\ldots,\prop_1,\prop_0) \big) }
            { \big( 1 - \adr_0(x,\prop_0) \big) \big( 1 - \adr_1(x,\prop_0,\prop_1) \big) \cdots \big( 1 - \adr_{j-1}(x,\prop_0,\prop_1,\ldots,\prop_{j-1}) \big) }
        \bigg) ~,
    \end{eqnarray}

    With every new rejection during the DR process, the proposal distribution can be reshaped to explore more probable regions of the parameter space. Therefore, the DRAM algorithm \citep{haario2006dram} combines the global adaptation capabilities of the AM algorithm \citep{haario2001adaptive} with the local adaptation capabilities of the DR algorithm \citep{green2001delayed}. In the following section, we present a variant of the generic DRAM algorithm that has been implemented in the ParaDRAM routine of the ParaMonte library.
    \newpar

%%%%%%%%%%%%%%%%%%%%%%%%%%%%%%%%%%%%%%%%%%%%%%%%%%%%%%%%%%%%%%%%%%%%%%%%%%%%%%%%%%%%%%%%%%%%%%%%%%%%%%%%%%%%%%%%%%%%%%%%%%%%%%%%%%%

\section{The ParaDRAM algorithm}
\label{sec:paradram}

    The DRAM algorithm of \S\ref{sec:dram} lays the foundation for the ParaDRAM algorithm. However, a major design goal of ParaDRAM is to provide a {\it fast} parallel Delayed-Rejection Adaptive Metropolis-Hastings Markov Chain Monte Carlo sampler. This is somewhat contrary to the complex generic form of the acceptance probability of the DRAM algorithm in \eqref{eq:acceptanceRatej} which can become computationally demanding for very high numbers of DR stages. Furthermore, a problem-specific adaptation strategy is generally required to successfully incorporate the knowledge acquired at each stage of the DR to construct the proposal for the next-stage DR. This is often a challenging task. We note, however, that \eqref{eq:acceptanceRatej} can be greatly simplified in the case of symmetric Metropolis proposals \citep{mira2001metropolis} where the probability of the proposed state only depends on the last rejected state,

    \begin{equation}
        \label{eq:dramSymmetricProposal}
        \pdr_{j}(\prop_j|\prop_{j-1},\ldots,\prop_0,x)=\pdr_{j}(\prop_j|\prop_{j-1})=\pdr_{j}(\prop_{j-1}|\prop_{j}) ~.
    \end{equation}

    In such case, the DR acceptance probability becomes,

    \begin{equation}
        \label{eq:acceptanceRateSymmetric}
        \adr_j(x,\prop_0,\prop_1,\ldots,\prop_j) = \min \bigg( 1 , \frac
        {
            \max \big( 0 , f(\prop_j) - f(y^*) \big)
        }{
            f(x) - f(y^*)
        }
        \bigg) ~~,~~ y^* = \argmax_{0\leq k<j} f(\prop_k) ,~j>0 ~.
    \end{equation}

    This symmetric version of the DRAM algorithm provides a fair compromise between the computational efficiency and the variance-reduction benefits of delaying the rejection. Therefore, we implement the symmetric Metropolis version of the DRAM algorithm in ParaDRAM and defer the implementation of the generic asymmetric form of DRAM to future work. \newpar

    Despite the symmetry of the algorithm described above, we note that the proposal corresponding to each stage of the DR can still be completely independent of the proposals at other stages of the DR or the zeroth-stage (i.e., the Adaptive algorithm's) proposal distribution. Ideally, the proposal kernel at each stage should be constructed by incorporating the knowledge of the rejected states in the previous stages. In practice, however, such informed proposal construction is challenging.
    \newpar

    By contrast, fixing the proposal distributions at all stages of the DR will be frequently detrimental to the performance of the sampler, since delaying the rejection often leads to steps that take the sampler into the vast highly-unlikely valleys and landscapes in the domain of the objective function that yield extremely small acceptance probabilities. In absence of any other relevant information about the structure of $f(x)$, a fair compromise can be made again by allowing the scale factor of the proposal distribution of each level of the DR to either shrink gradually from one stage to the next or be specified by the user before the simulation. The DR process can be then stopped either after a fixed number of stages (again, specified by the user) if all previous DR-stages have been unsuccessful, or by flipping a coin at each DR stage to continue or to stop and return to the AM algorithm. The former strategy is what we have implemented in ParaDRAM. The continuous process of adaptation of the proposal distribution as well as the DR process that is implemented in ParaDRAM is described in Algorithm \ref{alg:dram}.

    \begin{algorithm}[t]
    	\SetAlgoLined
        \caption{The symmetric-proposal DRAM algorithm as implemented in ParaDRAM \label{alg:dram}}
        \hrulefill\\
        \code{\textbf{Input : }} The natural logarithm of the target objective density function, $f(x)$. \\
    	\code{\textbf{Input : }} An initial starting point, $x_0$, for the DRAM pseudo-Markov chain. \\
        \code{\textbf{Input : }} An initial proposal distribution with PDF $\pam_0(\cdot)$. \\
        \code{\textbf{Input : }} The desired number of states, $N$, to be sampled from $f(x)$. \\
        \code{\textbf{Input : }} The maximum possible number of delayed-rejection stages at each MCMC step, $M$. \\
        \code{\textbf{Input : }} The vector $S=\{s_1,\ldots,s_M\}$ containing the scale factors of the DR proposal distributions. \\
        \code{\textbf{Output: }} The pseudo-Markov chain $x_1, \dots, x_N$\\
    	\textit{Initialize}\\
    	\For{i = 1 to $N$} {
            \begin{enumerate}
                \item
                    Propose a candidate state $\Prop$ from the AM proposal distribution with probability $\pam_{i-1}( \Prop=\prop | x_{i-1}, \dots, x_1, x_0 )$. \\
                \item
                    Set $x_i = \prop$ with probability, \\
                    \begin{equation}
                        \accr(x_{i-1},\prop) = \min\bigg( 1 , \frac{f(\prop)}{f(x_{i-1})} \bigg) ~, \nonumber
                    \end{equation}
                    \noindent otherwise, set $\pdr_0 = \pam_{i-1}$, $\prop_0 = \prop$, $\adr_0(\cdot,\cdot) = \accr_{i-1}(\cdot,\cdot)$ \\
            		\For{j = 1 to $M$} {
            			\begin{enumerate}
                            \item
                                Construct the $j$th-stage delayed-rejection proposal distribution with PDF $\pdr_j(\cdot|\cdot)$, \\
                                by rescaling the $(j-1)$th-stage DR proposal with PDF $\pdr_{j-1}(\cdot|\cdot)$ \\
                                with the user-provided scale factor $s_j$.
                            \item
                                Propose a new candidate $\Prop_j=\prop_j$ with probability $\pdr_j( \prop_j | \prop_{j-1} )$. \\
                            \item
                                Set $x_i = \prop_j$ with probability, \\
                                \begin{equation}
                                    \label{eq:acceptanceRateSymmetricAlgorithm}
                                    \adr_j(x,\prop_0,\prop_1,\ldots,\prop_j) = \min \bigg( 1 , \frac
                                    {
                                        \max \big( 0 , f(\prop_j) - f(y^*) \big)
                                    }{
                                        f(x) - f(y^*)
                                    }
                                    \bigg) ~, \nonumber
                                \end{equation}
                                where,
                                \begin{equation}
                                    y^* = \argmax_{0\leq k<j} f(\prop_k) ,~j>0 ~. \nonumber
                                \end{equation}
                                \noindent and \textbf{break for}, \\
                                otherwise, \textbf{continue}
                        \end{enumerate}
                    }
                    \uIf{$j>M$}{
        				$x_i \gets x_{i-1}$, \\
                        \textbf{continue} (no candidate was accepted in the delayed rejection stages)
        			}
    	   \end{enumerate}
        }
    \end{algorithm}

    \subsection{Ensuring the diminishing-adaptation of the DRAM algorithm}
    \label{sec:paradram:diminishing}

        In practice, the DRAM algorithm has to stop after a finite number of iterations, for example, as specified by $N$ in Algorithm \ref{alg:dram}. Since the convergence and ergodicity of the chain generated by the DRAM algorithm is valid only asymptotically, it is crucial to monitor and ensure the asymptotic convergence of the chain to the target probability density function which, with a slight abuse of notations, we represent by $f(x)$. The convergence can be ensured by measuring the {\it total variation} between the target and the generated distribution from the $n$th-stage adapted proposal,

        \begin{equation}
            \lim_{n\rightarrow+\infty} \norm{ \pi_n(\cdot|x) - f(\cdot) } = 0 ~.
        \end{equation}

        This is, however, impossible since the sole source of information about the shape of the target density is the generated chain. To resolve this problem of non-ergodicity of the finite chain due to the continuous adaptation, one conservative community approach has been to perform the adaptation for only a limited time. After a certain period, the adaptation fully stops and the regular MH-MCMC simulation begins with a fixed proposal. Consequently, the entire chain before fixing the proposal is thrown away and the final refined samples are generated only from the fixed-proposal MCMC chain.
        \newpar

        This sampling approach, known as the {\it finite adaptation} \citep{kass1998markov, roberts2007coupling} is essentially identical to the traditional MH-MCMC approach except in the initial automatic fine-tuning of the proposal distribution, which is done manually in the traditional MH-MCMC algorithm. Although the finite adaptation approach ensures the ergodicity and the Markovian properties of the resulting chain, it suffers from the same class of limitations of the MH-MCMC algorithm. For example, it is not clear when the adaptation process should stop, and what should be the criterion used to automatically determine the stopping time of the adaptation. If the adaptation stops too early in the simulation before good-mixing occurs, the resulting fixed-proposal MH-MCMC simulation can potentially suffer from the same slow-convergence issues encountered with the use of MH-MCMC algorithm, necessitating a restart of the adaptive phase of the simulation. This process of adaptation and verification will have to be then continued for as long as needed until the user can confidently fix the proposal distribution to generate the final Markovian chain.
        \newpar

        Here we propose a workaround for this problem by noting that the entire adaptation in the DRAM algorithm is contained within the proposal distribution, $\pam(\cdot)$. Therefore, if we can somehow measure the amount of change between the subsequent adaptations of the proposal distribution, we can indirectly and dynamically assess the importance and the total effects of the adaptation on the chain that is being generated in real-time.
        \newpar

        One of the strongest measures of the difference between two probability distributions is given by the metric {\it total variation distance (TVD)} between the two. For the two distributions, $\Pam_i$ and $\Pam_{i+1}$, defined over the $d$-dimensional space $\mathbb{R}^{d}$ with the corresponding densities, $\pam_i$ and $\pam_{i+1}$, the TVD is defined as \citep{tsybakov2008introduction},
        \begin{equation}
            \label{eq:tvd}
            {\mathrm{TVD}}(\Pam_i,\Pam_{i+1}) = \frac{1}{2} \int_{\mathbb{R}^{d}} \big| \pam_i(x) - \pam_{i+1}(x) \big| ~ \diff x ~.
        \end{equation}

        In other words, TVD is half of the $L^1$ distance between the two distributions. The TVD is by definition a real number between $0$ and $1$, with $0$ indicating the identity of the two distributions and, $1$ indicating two completely different distributions. Despite its simple definition, the computation of TVD is almost always intractable, effectively rendering it useless in our practical ParaDRAM algorithm.
        \newpar

        To overcome the difficulties with the efficient computation of the TVD, we substitute the TVD with an upper bound on the its value. By definition, this upper bound always holds for any arbitrary pair of distributions and is defined via another metric distance between the two probability distributions, known as the Hellinger distance \citep{hellinger1909neue}, whose square is defined as,
        \begin{equation}
            \label{eq:hellinger}
            {\mathrm{H}}^2 (\Pam_i,\Pam_{i+1}) = 1 - \int_{\mathbb{R}^{d}} \sqrt{ \pam_i(x)~ \pam_{i+1}(x) } ~ \diff x ~.
        \end{equation}

        By definition, the Hellinger distance is always bounded between 0 and 1, with 0 indicating the identity of the two distributions and, 1 indicating completely different distributions. A simple reorganization of the above equation reveals that the Hellinger distance is the $L^2$ distance between $\sqrt{\pam_{i}}$ and $\sqrt{\pam_{i+1}}$. Furthermore, it can be shown that the following inequalities hold between the Hellinger distance and the TVD \citep{tsybakov2008introduction},
        \begin{equation}
            \label{eq:hellingerTVD}
            \frac{1}{2} {\mathrm{H}}^2 (\Pam_i,\Pam_{i+1}) \leq
            {\mathrm{TVD}}(\Pam_i,\Pam_{i+1}) \leq
            {\mathrm{H}}(\Pam_i,\Pam_{i+1}) \sqrt{1 - \frac{{\mathrm{H}}^2(\Pam_i,\Pam_{i+1})}{4} } \leq
            {\mathrm{H}}(\Pam_i,\Pam_{i+1}) ~.
        \end{equation}

        Unlike TVD, the computation of the Hellinger distance is generally more tractable. In particular, the Hellinger distance has closed-form expression in the case of the MultiVariate Normal (MVN) distribution, which is, by far, the most widely-used proposal distribution in all variants of the MCMC method, including the DRAM algorithm. Even in cases where a closed-form expression for a proposal distribution may not exist, the TVD upper-bound computed under the assumption of an MVN proposal distribution might still provide an upper-bound for the TVD of the proposal distribution of interest, under some conditions.
        \newpar

        Therefore, we use the inequalities expressed in \eqref{eq:hellingerTVD} to estimate an upper bound for total variation distance between any two subsequent updates of the proposal distribution in the DRAM algorithm. This enables us to dynamically monitor and ensure the diminishing adaptation of the DRAM simulations. In practice, we find that the progressive amount of the adaptation of the proposal distribution diminishes fast as a power-law in terms of the MCMC steps. In cases of rapid good mixing, the initial few thousands of steps of the simulation exhibit significant adaptation of the proposal, followed by a fast power-law drop in the amount of adaptation. Some examples of the dynamic adaptation monitoring of the DRAM simulations are discussed in \S\ref{sec:results}.
        \newpar

        This practical method of dynamically measuring the adaptation also fulfills one of the major conditions for the ergodicity of the DRAM algorithm, for as long as the simulation can continue \citep[e.g., see theorem (1) in][]{roberts2007coupling}. It also provides an indirect qualitative method of rejecting the convergence to the target density if the TVD upper-bound estimate fails to continuously decrease or, even further increases with the progression of the DRAM simulation.

    \subsection{The parallelization of the DRAM algorithm}
    \label{sec:paradram:parallelism}

        Contemporary scientific problems typically require parallelism to obtain solutions within a reasonable time-frame. As such, a major cornerstone of the ParaDRAM algorithm and the ParaMonte library is to enable seamless parallelization of Monte Carlo simulations without requiring any parallel programming experience from the user. Furthermore, to ensure the scalability of the ParaDRAM algorithm, from personal laptops to hundreds of cores on supercomputers, we have intentionally avoided the use of shared-memory parallelism in the algorithm, most notably, via the OpenMP \citep{dagum1998openmp} or OpenACC \citep{enterprisecray} standards. Nevertheless, this mode of parallelism remains a viable choice for future work.
        \newpar

        Instead, the entire parallelization of the ParaDRAM algorithm is currently done via two independent scalable {\it distributed-memory} parallelism paradigms: 1. the {\it Message Passing Interface standard (MPI)} \citep{gropp1996high} and, 2. the {\it Partitioned global address space (PGAS)} \citep{numrich1998co, fanfarillo2014opencoarrays}. Unlike shared-memory parallelism, the distributed-memory architecture allows for scalable simulations beyond a single node of physical processors, across a network of hundreds or possibly, thousands of processors. This is an essential feature for parallel Monte Carlo algorithms in the era of ExaScale computing \citep{bergman2008exascale}, although we will discuss some limitations of the current parallelism implementation of ParaDRAM in \S\ref{sec:discussion}.
        \newpar

        The PGAS parallelism paradigm readily enables Remote Memory Access (RMA), commonly known as one-sided communication, from one processor to all other processors in parallel simulations. This allows multiple data transfers between a set of processes to use a single synchronization operation, thus reducing the total overhead of inter-processor communications. By contrast, the RMA communications via MPI are considerably more challenging to implement. Nevertheless, the current support for the MPI parallelism paradigm is more robust than for the PGAS paradigm. Consequently, the utility of the PGAS parallelization of ParaDRAM currently remains limited to the Fortran language interface to the ParaDRAM algorithm, enabled by the Coarray Fortran \citep{metcalf2018modern}. Conversely, the MPI-parallelized version of ParaDRAM is accessible from all available programming language interfaces to ParaDRAM (e.g., C, C++, Fortran, MATLAB, Python, ...), even where the programming language does not officially support the MPI paradigm.
        \newpar

        For both the MPI and PGAS communication paradigms in ParaDRAM, two parallelization models are currently implemented: 1. The Fork-Join parallelism \citep{conway1963multiprocessor} and, 2. The Perfect parallelism \citep{moler1986matrix}.

        \begin{figure}[t!]
            \centering
            \begin{subfigure}[t]{\textwidth}
                \centering
                \includegraphics[width=\textwidth]{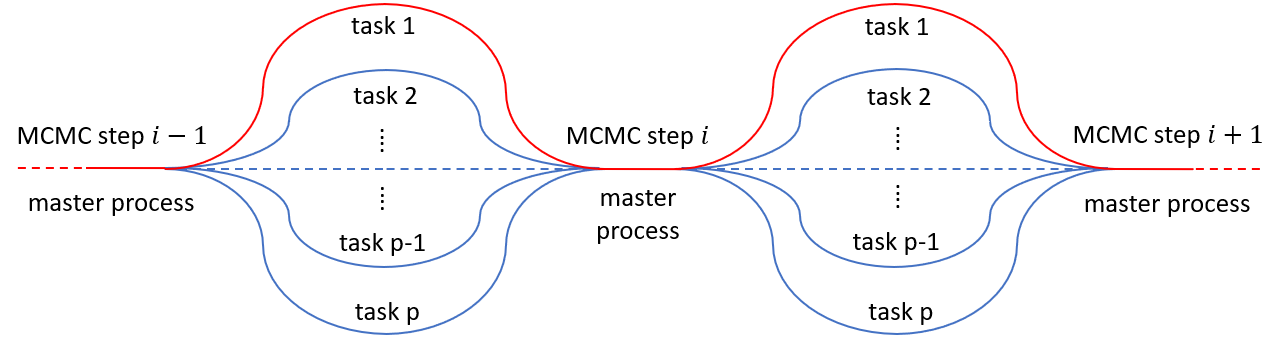}
                \caption{The fork-join parallelism.} \label{fig:forkJoinParallelism}
            \end{subfigure}
            \par\bigskip % force a bit of vertical whitespace
            \begin{subfigure}[t]{\textwidth}
                \centering
                \includegraphics[width=\textwidth]{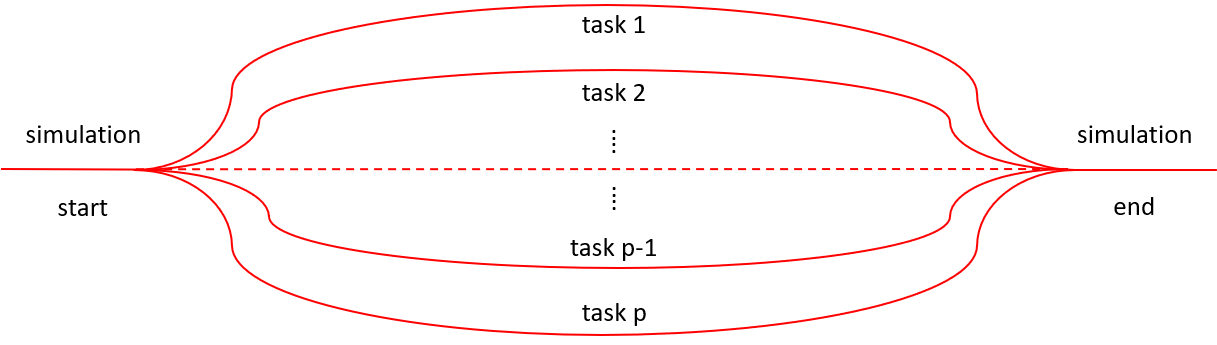}
                \caption{The prefect parallelism.} \label{fig:perfectParallelism}
            \end{subfigure}
            \caption{{\bf(a)} An illustration of the fork-join parallelism implemented in the current version of the ParaDRAM algorithm. At each iteration of the MCMC simulation, a master process (represented by the {\textcolor{red}{red-line}}) distributes the current state of the sampler with all other processes (represented by the {\textcolor{blue}{blue-lines}}). Then each process proposes a move which is either accepted or rejected and the result is returned to the master process for a final decision. {\bf(b)} An illustration of the perfect parallelism implemented in the current version of the ParaDRAM algorithm. Each process runs an MCMC simulation independently of the rest of the processes. The communication cost is, therefore, zero in perfect parallelism.}
        \end{figure}

        \subsubsection{The Fork-Join parallelism}
        \label{sec:paradram:parallelism:forkjoin}

            In this (\texttt{single-chain}) parallelism mode, the zeroth MPI process (or the first Coarray image) in the simulation is the master process responsible for reading the input data, updating the proposal distribution of the DRAM algorithm, concluding the simulation, and performing any subsequent post-processing of the simulation output data. All other processes in the simulation communicate and share information only with the master process/image \ref{fig:forkJoinParallelism}.
            \newpar

            At each MCMC iteration, information about the current step is broadcasted by the master process to all other processes. Then, each process, including the master, proposes a new state for the chain and calls the user-provided objective function independently of the other processes. The proposed states together with the corresponding objective function values are then communicated to the master process. The master process then checks the occurrence of an accepted new state, proposal by any of the processes including itself. Upon the occurrence of the first acceptance, it communicates the new accepted state to all other processes and the next MCMC step begins. If no acceptance occurs, either the same old state is communicated to all processes to continue with the next MCMC step or, the simulation enters the Delayed-Rejection (DR) phase if requested by the user.
            \newpar

            When the DR is enabled, the simulation workflow is similar to the above, except that upon rejecting the proposed moves by all processes at each stage of the DR, each process is allowed to only and only take one more delayed-rejection step. The results are then sent back to the master process to decide on whether the next DR stage has to be initiated. However, if an acceptance occurs on any of the processes at any given DR stage, the DR algorithm stops and the simulation returns to the regular adaptive algorithm. However, if no acceptance occurs after the maximum number of DR stages has reached, the last accepted state is communicated again to all processes, and the workflow of adaptive sampling repeats.
            \newpar

            This mode of communication between the processes at each stage of the DR is crucial for the computational efficiency of the ParaDRAM algorithm since it is often significantly more costly to call the objective function redundantly (until the maximum number of DR stages is reached) than to communicate a few bytes of information between the processes at each DR stage to check if any acceptance has occurred. In \S\ref{sec:results}, we will present some performance benchmarking results for the two MPI and PGAS fork-join parallelism implementations in ParaDRAM. 

        \subsubsection{The Perfect parallelism}
        \label{sec:paradram:parallelism:perfect}

            In the Perfect (\texttt{multi-chain}) parallelism mode, each MPI process (or Coarray image) runs independently of the other processes (images) to create its DRAM pseudo-Markov chain. This is effectively equivalent to having as many serial versions of the ParaDRAM algorithm to run concurrently and simulate independently of each other. However, unlike multiple independent concurrently-run serial DRAM chains, the ParaDRAM algorithm in the perfect-parallelism mode also performs post-simulation pairwise comparisons of the resulting chains from all processes to check for the convergence of all chains to the same unique target density.
            \newpar

            To ensure the multi-chain convergence to the same target density, first, the initial burnin episodes of all chains are automatically removed and each chain is iteratively and aggressively refined (i.e., decorrelated) to obtain a final fully-decorrelated Independent and Identically Distributed (\iid) sample from the target density, corresponding to each chain. Then, the similarities of the individual corresponding columns of all chains are compared with each other via the two-sample Kolmogorov-Smirnov (KS) nonparametric test \citep{kolmogorov1933sulla, smirnov1948table}. Finally, the results of the tests are reported in the output {\it report} file corresponding to each generated chain.
            \newpar

            The perfect \texttt{multi-chain} parallelism, as explained in the above, has the benefit of providing an automatic convergence-checking, via the KS test, at the end of the simulation. This is a remarkable benefit that is missing in the fork-join \texttt{single-chain} parallelism. On the flip side, the perfect parallelism quickly becomes inferior to the fork-join paradigm for large-scale MCMC simulations, since the pressing issue in such cases is the computational and runtime efficiency of the simulation.

    \subsection{The final sample refinement}
    \label{sec:paradram:refinement}

        A unique feature of the Markov Chain Monte Carlo simulations is the Markovian property of the resulting chain, which states that the next step in the simulation given the currently visited state is independent of the chain's past. It may, therefore, sound counterintuitive to realize that the resulting sample from a finite-size MCMC simulation could be highly autocorrelated. Notably, however, each visited state in the chain depends on the last state visited before it. This implicit sequential dependence of all points on their past, up to the starting point, is what creates significant autocorrelations within MCMC samples.
        \newpar

        For the infinite-length Markov chain of \eqref{eq:randomSeq} that has converged to its stationary equilibrium distribution, the autocorrelation function is defined as,
        \begin{equation}
            \label{eq:randomSeqAC}
            \acf(k) = \frac{ \mathbb{E} \big[ (X_i - \mu) (X_{i+k} - \mu) \big] } { \sigma^2 } ~,
        \end{equation}

        \noindent where $(\mu,\acf(0)=\sigma^2)$ represent the mean and the standard deviation of the Markov chain and $\mathbb{E}[\cdot]$ represents the expectation operator. The {\it Integrated Autocorrelation (IAC)} of the chain is defined with respect to the variance of the estimator of the mean value $\mu$,
        \begin{equation}
            \label{eq:randomSeqIAC}
            \iac = 1 + 2\sum_{k=1}^{+\infty} \acf(k) ~,
        \end{equation}

        \noindent such that,
        \begin{equation}
            \label{eq:randomSeqLimitIAC}
            \lim_{n\rightarrow+\infty} \sqrt{\frac{n}{\iac}} \frac{\mu_n - \mu}{\sigma} \Rightarrow N(0,1) ~,
        \end{equation}

        \noindent where $\mu_n$ represents the sample mean of the chain of length $n$ and `$\Rightarrow$' stands for convergence in distribution. The value of $\iac$ roughly indicates the number of Markov transitions required to obtain an \iid~sample from the target distribution of the Markov chain. In practice, one wishes to obtain a finite MCMC sample whose size is at least of the order of the integrated autocorrelation of the chain \citep{robert2010introducing}. This is a challenging goal that is often out of reach since the \iac~of the Markov chain is not known a priori. A more accessible approach is to generate a chain for a given predefined length, then de-correlate it to obtain a final {\it refined} independent and identically distributed (i.i.d.) sample from the target density.
        \newpar

        The numerical computation of the \iac, however, poses another challenge to decorrelating MCMC samples since the variance of its estimator in \eqref{eq:randomSeqIAC} diverges to infinity. A wide variety of techniques have been proposed that aim to estimate the \iac~for the purpose of MCMC sample refinement. Among the most popular methods are the Batch Means (BM) \citep{fishman1978principles}, the Overlapping Batch Means (OBM) \citep{schmeiser1982batch}, the spectrum fit method \citep{heidelberger1981spectral}, the initial positive sequence estimator \citep{geyer1992practical}, as well as the auto-regressive processes \citep[e.g.,][]{plummer2006coda}.
        \newpar

        \citet{thompson2010comparison} performs a series of tests aimed at identifying the fastest and the most accurate method of estimating the \iac. They find that while the auto-regressive process appears to be the most accurate method of estimating the \iac, the Batch Means method provides a fair balance between the computational efficiency and numerical accuracy of the estimate.
        \newpar

        Based on the findings of \citet{thompson2010comparison}, we have therefore implemented the Batch Means method as the default method of estimating the \iac~of the resulting Markov chains from the ParaDRAM sampler. Notably, however, all the aforementioned methods appear to underestimate the value of \iac, in particular, for small chain sizes. Therefore, we have adopted a default aggressive methodology in the ParaDRAM algorithm where the autocorrelation of the chain is removed repeatedly (via any estimator of choice by the ParaDRAM user, such as the BM) until the final repeatedly-refined chain does not exhibit any autocorrelation.
        \newpar

        This aggressive refinement of the chain is performed in two separate stages: At the first stage, the full Markov chain is repeatedly refined based on the estimated \iac~values from the (non-Markovian) {\it compact} chain of the uniquely accepted points. This stage essentially removes any autocorrelation in the Markov chain that is due to the choice of too-small step sizes for the proposal distribution (Figure \ref{fig:smallStepSampling}). Once the compact chain of accepted points is devoid of any autocorrelations, the second phase of the Markov chain refinement begins, with the \iac~values now being computed from the ({\it verbose}) Markov chain, starting with the resulting refined Markov chain from the first stage of the refinement (of the compact chain).
        \newpar

        We have found by numerous experimentations that the above approach often leads to final refined MCMC samples that are fully decorrelated while not being refined too much due to our aggressive repetitive decorrelation of the full Markov chain. Nevertheless, the above complex methodology for the refinement of the Markov chain can be entirely controlled by the input specifications of the simulation set by the user. For example, the user can request only one round of chain refinement to be performed using only one of format of the chain: compact or verbose (Markov).
        \newpar

\section{One API for ParaDRAM across all programming languages}
\label{sec:api}

    Special care has been made to ensure that the Application Programming Interface (API) of the ParaDRAM algorithm retains highly-similar (if not the same) structure, naming, and calling conventions across all programming languages currently supported by the ParaMonte library. First and foremost, the interface to the ParaDRAM routine requires only two mandatory pieces of information to be provided by the user:

    \begin{enumerate}
        \item
            {\bf \texttt{ndim}}: the dimension of the domain of the objective function to be sampled and,
        \item
            {\bf \texttt{getLogFunc(ndim,point)}}: a computational implementation of the objection function in the programming language of choice, which should take as input a 32-bit integer \texttt{ndim} and a 64-bit real vector \texttt{point} of length \texttt{ndim} that represents a state from within the domain of the objective function. On return, the function yields the natural logarithm of the value of the objective function evaluated at \texttt{point}. Unlike C/C++/Fortran, in the case of higher-level programming languages such as MATLAB or Python, the calling syntax of the objective function simplifies to \texttt{getLogFunc(point)} where the length of \texttt{point} is passed implicitly.
    \end{enumerate}

    The one-API paradigm has been one of the core design philosophies of the ParaMonte library (including the ParaDRAM algorithm) to ensure similar user experience and the availability of the same functionalities from all supported programming language interfaces to the ParaMonte / ParaDRAM library. The full description of all capabilities and details of each of the programming-language interfaces to the ParaDRAM routine goes well beyond the limitations of the current manuscript. Therefore, we will only present some of key identical components of the algorithm shared among all available interfaces to the ParaDRAM sampler.

    \subsection{The ParaDRAM simulation specifications}
    \label{sec:api:input}

        The ParaDRAM sampler has been mindfully developed to be as flexible as possible regarding the settings of the simulations. Consequently, there is a long list of input simulation specifications whose complete descriptions go beyond the limits of this paper. We refer the interested reader to permanent repository\footnote{\url{https://github.com/cdslaborg/paramonte}} and the documentation website\footnote{\url{https://www.cdslab.org/paramonte/}} of the ParaMonte library for the detailed descriptions of the simulation specifications.
        \newpar

        Despite the great number and variety of the ParaDRAM simulation specifications, all 39 independent input specification variables currently available in ParaDRAM are optional and automatically set if not provided by the user. In some simulation scenarios, some level of input information may be necessary from the user, for example, when the domain of the objective function does not extend to infinity, in which case, the user can readily specify the boundaries of a cube within which the sampling will be performed.
        \newpar

        From within the compiled programming languages, the preferred method of specifying simulation parameters is to store them all within an input file and provide the path to this file at the time of calling the ParaDRAM routine. This approach enables changes to the simulation configuration seamlessly possible without any need to recompile and relink the source codes to build a new executable, which can be a time-consuming process for large-scale simulations. From within all compiled languages (C/C++/Fortran), the simulation specifications can be also passed as a string to the ParaDRAM sampler, instead of passing the path to an external file. In such case, the value of the string could be the contents of the input file (instead of the path to the input file). From within the Fortran language, the users can also pass the simulation specifications as \texttt{optional} input arguments to the ParaDRAM sampler.
        \newpar

        From within the interpreted programming languages such as MATLAB and Python, the preferred method of specifying the simulation configuration is via the dedicated Object-Oriented Programming (OOP) interface that we have developed in each of these programming language environments. Nevertheless, the users can also alternatively provide the same input file that is used in compiled languages to configure their ParaDRAM simulations. Given the great flexibility of the interpreted languages, specifying the simulation configuration via an input file seems to be inferior to the OOP interface that we have written for each of these programming language environments.
        \newpar

    \subsection{The ParaDRAM simulation output files}
    \label{sec:api:output}

        Each ParaDRAM simulation, performed from within any programming language environment, generates five distinct output files. If the user has specified a simulation name then all output files {\it prefixed} are prefixed by the user-provided simulation name. Otherwise, the output files are prefixed by a unique automatically-generated random simulation name with a specific pattern, for example: \texttt{"./out/ParaDRAM\_run\_20200101\_205458\_278\_process\_1"}, where,

        \begin{enumerate}
            \item
                \texttt{./out/} is the example user-requested directory within which the output simulation files are stored (and if the specified directory does not exist, it is automatically generated),
            \item
                \texttt{ParaDRAM} indicates the type of the simulation,
            \item
                \texttt{run\_yyyymmdd\_hhmmss\_mmm} indicates the date of the simulation specified by the current year (\texttt{yyyy}), month (\texttt{mm}), and day (\texttt{dd}), followed by the start time of the simulation specified by the hour (\texttt{hh}), the minute (\texttt{mm}), the second (\texttt{mm}), and the millisecond (\texttt{mmm}) of the moment of the start of the simulation,
            \item
                \texttt{process\_1} indicates the ID of the processor that has generated the output files, with \texttt{1} indicating the master process (or Coarray image).
        \end{enumerate}

        The above random prefix-naming convention both documents the exact date and time of the simulation and ensures the uniqueness of the names of the generated output files. In the extremely-rare event of a user-specified filename clash with an existing set of simulation files in the same directory, the simulation will be aborted and the user will be asked to specify a unique name for the new simulation.
        \newpar

        Once the uniqueness of the prefix of the simulation output files is ensured, the ParaDRAM sampler generates five separate output files with the same prefix, but with the following suffixes that imply the purpose and the type of the contents of each file,

        \begin{enumerate}
            \item
                \texttt{\_chain.txt} or \texttt{\_chain.bin} indicates a file containing the ParaDRAM output Markov Chain, where the user's choice of the format of the file (\texttt{ASCII} vs. \texttt{binary}) dictates the file extension (\texttt{txt} vs. \texttt{bin}).
            \item
                \texttt{\_sample.txt} indicates a file containing the final refined decorrelated sample from the target density, containing only the refined set of visited states and their corresponding target density values reported in natural logarithm.
            \item
                \texttt{\_report.txt}: indicates a file containing a comprehensive report of all aspects of the simulation, including the ParaDRAM library version, the computing platform, the user-specified description of the simulation, the user-specified (or automatically-determined) simulation configuration, the description of the individual simulation specifications, any runtime simulation warnings or fatal errors, as well as extensive report on the timing and performance of serial/parallel simulation and extensive postprocessing of the simulation results.
            \item
                \texttt{\_progress.txt}: indicates a file containing a dynamic report of the simulation progress, such as the dynamic efficiency of the MCMC sampler, time spent since the beginning of the simulation and the predicted time remained to accomplish the simulation.
            \item
                \texttt{\_restart.txt} or \texttt{\_restart.bin}: indicates a file containing information required for a deterministic restart of the simulation, should the simulation end prematurely. The user's choice of the format of the file (\texttt{ASCII} vs. \texttt{binary} dictates the file's extension (\texttt{txt} vs. \texttt{bin}).
        \end{enumerate}

    \subsection{Efficient compact storage of the Markov Chain}
    \label{sec:api:storage}

        The restart functionality and the ability to handle large-scale simulations that exceed the random-access-memory (RAM) limits of the processor require the ParaDRAM algorithm to continuously store the resulting chain of sampled points throughout the simulation. However, this poses two major challenges to the high efficiency and low memory-footprint of the algorithm:

        \begin{enumerate}
            \item
               Given the current computational technologies, input/output (IO) from/to external files is on average 2-4 orders of magnitude slower than the RAM IO. This creates a severe bottleneck in the speed of the otherwise high-performing ParaDRAM algorithm, in particular, for large-scale high-dimensional objective functions.
            \item
                Moreover, the resulting output chain files can easily grow to several Gigabytes, even for regular MCMC simulations, making the storage of multiple simulation output files over the long term challenging or impossible.
        \end{enumerate}

        To minimize the effects of external IO on the performance and the memory-footprint of the algorithm, we propose to store the resulting chain of states from ParaDRAM in a very compact format that dramatically enhances the library's performance and lowers its memory footprint 5-10 times, without compromising the fully-deterministic restart functionality of ParaDRAM or its ability to handle large-scale memory-demanding simulations.
        \newpar

        The {\it compact} (as opposed to {\it verbose} or Markov) storage of the chain is made possible by noting that the majority of states in a typical Markov chain are identical as a result of the repeated rejections during the sampling. The lower the acceptance probability is, the larger the fraction of repeated states in the verbose Markov chain will be. Therefore, the storage requirements of the chain can be dramatically reduced by keeping track of only the accepted states and assigning weights to them based on the number of times they are sequentially repeated in the Markov chain.
        \newpar

        Furthermore, since the contents of the output chain file is peripheral to the contents of the final refined sample file, the ParaDRAM sampler also provides a \texttt{binary} output mode, where the chain will be written out in binary format. Although the resulting output chain file is unreadable by human, the binary IO is fast, does not suffer from loss of precision due to the conversion from binary to decimal for external IO, and in general, occupies less memory for same level of accuracy. Nevertheless, we believe the above proposed \texttt{compact} chain file format provides a good compromise between, IO speed, memory-footprint, and readability. Therefore, we use the compact ASCII file as the default format of the output chains from ParaDRAM simulations. Users can also specify a third \texttt{verbose} format where the resulting Markov chain will be written to the output file {\it as is}. However, this \texttt{verbose} mode of chain IO is not recommended except for debugging or exploration purposes since it significantly degrades the algorithm's performance and increases the memory requirements for the output files.

    \subsection{The ParaDRAM simulation restart functionality}
    \label{sec:api:restart}

        An integral part of the ParaDRAM algorithm is its automatically-enabled fully-deterministic reproducibility of the simulation results, should a ParaDRAM simulation, whether serial or parallel end prematurely. In such cases, all the user needs to do in order to restart the simulation from where it was interrupted, is to rerun the simulation (with the same output file prefix).
        \newpar

        The ParaDRAM algorithm has been designed to automatically detect the existence of the output simulation files. If all the simulation files already exist, the simulation will be aborted with a message asking the user to provide a unique file-prefix name for the output simulation files. However, if all files exist except the output refined sample file, which is generated in the last stage of the simulation, ParaDRAM enters the restart mode and begins the simulation from where it was interrupted during the last run.
        \newpar

        A remarkable feature of the restart functionality of the ParaDRAM algorithm is its fully-deterministic reproduction of the simulation results {\it into the future}: If a simulation is interrupted and subsequently restarted, the resulting final chain after the restart would be identical, up to 16 digits of precision, to the chain that the sampler would have generated if the simulation had not been interrupted in the first place. To generate a fully deterministic reproducible simulation, all that is needed from the user is to set the seed of the random number generator of the ParaDRAM sampler as part of the input simulation specifications. Additionally, in the case of parallel simulations, it is expected that the same number of processes will be used to run the restart simulation as used for the original interrupted simulation.
        \newpar

        The information required for the restart of an interrupted ParaDRAM simulation is automatically written to the output restart file. To minimize the impacts of the restart IO on the performance and the external the memory requirements of the algorithm, the restart file is automatically written in binary format. This default behavior can be overridden by the user by requesting an \texttt{ASCII} restart file format in the input simulation specifications to the sampler. In such case, ParaDRAM will also automatically add additional relevant information about the dynamics of the proposal adaptations to the output file. This human-readable information can be then parsed to gain insight into the inner-workings of the ParaDRAM algorithm. An example of such analysis of the dynamics of the proposal adaptation will be later given in \S\ref{sec:results}.

    \subsection{The optimal number of processors for parallel ParaDRAM simulations}
    \label{sec:api:scaling}

        When a parallel MCMC simulation is performed in the fork-join (\texttt{single-chain}) parallelism mode, the ParaDRAM algorithm, as part of a default post-processing analysis, attempts to predict the {\it optimal} number of parallel processors from which the simulation could benefit. In general, the overall parallel efficiency of a software depends on a number of factors including (but not limited to),

        \begin{enumerate}
            \item $T_s$:
               The serial runtime required for all computations that cannot be parallelized and must be performed in serial mode.
            \item $T_p$:
                The serial runtime for the fraction of the code that can be equally shared among all processors in parallel.
            \item $T_o$:
                The time required for setting up the inter-processor communications, and information exchange, effectively known as the {\it communication overhead}.
        \end{enumerate}

        Among the three time measures mentioned in the above, the overhead time is the most complex and hardest to estimate since it is highly software and hardware dependent. Nevertheless, this overhead time can be frequently assumed to linearly grow with the number of processes in the parallel simulation. This is particularly true for the fork-join parallelism paradigm where all inter-process communications happen to and from a master process. Therefore, the overall speedup due to the use of $\nproc$ processors in parallel can be computed from a modified form of the Amdahl's law of strong scaling \citep{amdahl1967validity} that takes into account the communication overhead time,

        \begin{equation}
            \label{eq:maxSpeedupAbsolute}
            S(\nproc) \approx \frac
            {
                T_s + T_p ~ \text{(The total serial run time of the simulation)}
            }{
                T_s + \frac{T_p}{\nproc} + (\nproc-1) \times T_o
            }
            ~,
        \end{equation}

         In the case of the ParaDRAM routine, the runtime of the serial fraction ($T_s$) is typically on the order of a few tens of nanoseconds to microseconds on the modern architecture, whereas the parallel fraction of the simulation ($T_p$) -- which calls the user-provided objective function -- is expected to dominate the simulation runtime. Therefore, compared to $T_p$ and $T_o$, the serial fraction ($T_s$) can be safely ignored in large-scale ParaDRAM simulations. Then, to compute the speedup in any given parallel ParaDRAM simulation, $T_p$ and $T_o$ can be respectively estimated from the average runtimes of the parallel and the inter-process communication sections of the code.
        \newpar

        Once $T_p$ and $T_o$ are estimated, we can then predict the simulation speedup over a wide range of number of processors. The maximum predicted speedup then provides an {\it absolute} upper bound on the number of processors that could benefit the simulation. In practice, however, this {\it absolute optimal} number of processors is only an upper bound on the actual number of processors from which the given simulation would effectively benefit. In the special case of parallel fork-join MCMC simulations, there is yet another equally-important factor that, along with the communication overhead, limits the overall speedup of parallel simulations. This non-negligible factor is the efficiency of the MCMC sampler.
        \newpar

        The role of the average MCMC acceptance rate on the optimal number of processors can be understood by noting that the average number of MCMC steps that need to be taken before an acceptance occurs is roughly proportional to the inverse of the average MCMC acceptance rate. For example, if the average acceptance rate is $0.25$, then one would expect an acceptance to occur every four steps. This places a fundamental limit on the number of processors from which the simulation could benefit in parallel.
        \newpar

        Quantitatively, the process of accepting a proposed state in a given step of the ParaDRAM algorithm, parallelized over an infinite number of processors ($\nproc\rightarrow+\infty$), can be modeled as a Bernoulli trial with two possible outcomes: rejection or acceptance of the proposed state. In this process, the probability of an acceptance can be assumed to be represented by the average MCMC acceptance rate ($\maccr$). Thus, the probability of an acceptance occurring after $k$ proposals (by the first $k$ processors) follows a Geometric distribution, $\geodist(\cdot)$, and is given by,

        \begin{equation}
            \label{eq:bernoulli}
            \geoprob \big( \text{acceptance} ~|~ k \big) = \maccr ~ (1-\maccr)^{(k-1)} ~.
        \end{equation}

\begin{figure}[t!]
            \centering
            \begin{subfigure}[t]{0.48\textwidth}
                \centering
                \includegraphics[width=\textwidth]{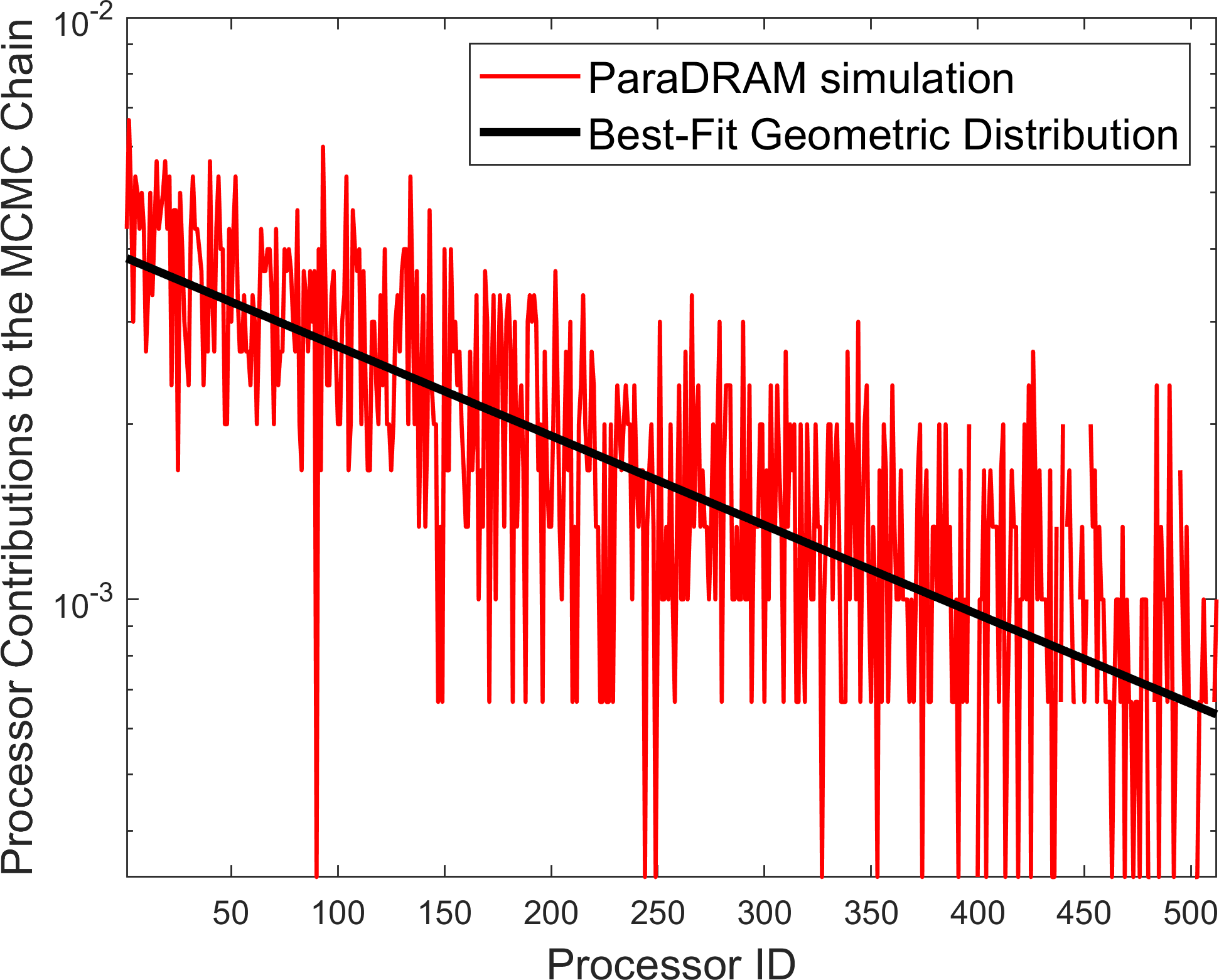}
                \caption{Processor contributions to a parallel simulation.} \label{fig:procContribution}
            \end{subfigure}
            \hfill
            \begin{subfigure}[t]{0.48\textwidth}
                \centering
                \includegraphics[width=\textwidth]{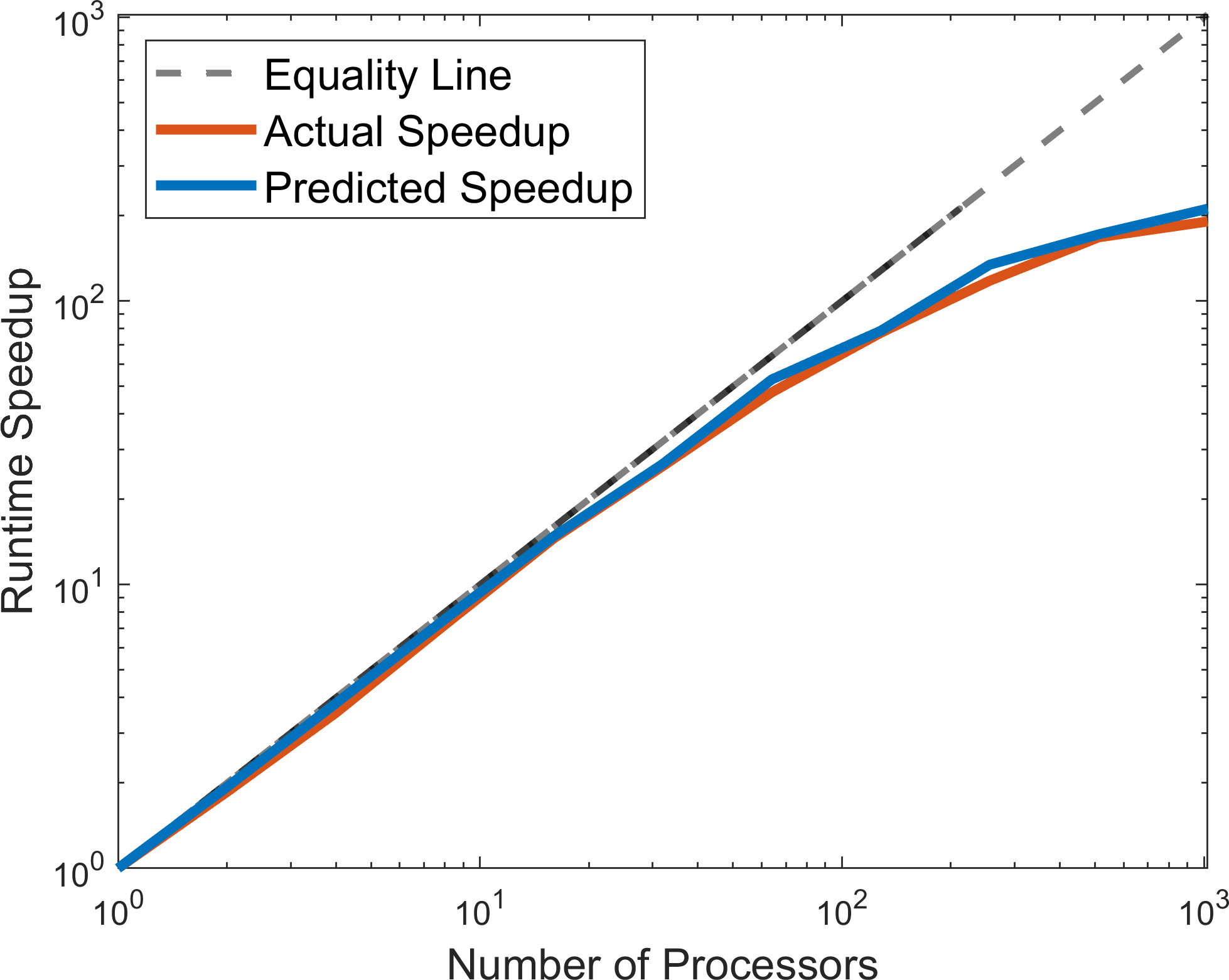}
                \caption{The parallel strong-scaling: predicted vs. actual.} \label{fig:PredictedSpeedupActualSpeedup}
            \end{subfigure}
            \caption{{\bf(a)} An illustration of the contributions of 512 Intel Xeon Phi 7250 processors to a ParaDRAM simulation in parallel (the \textcolor{red}{red curve}). The predicted best-fit Geometric distribution from the post-processing phase of the ParaDRAM simulation is shown by the black line. {\bf(b)} A comparison of the parallel-performance of ParaDRAM simulations on a range of processor counts with the performance predictions from the post-processing output of the ParaDRAM sampler. The entire performance data depicted in the plots (a) and (b) of this figure are automatically generated by the ParaDRAM sampler as part of the post-processing of every parallel MCMC simulation.}
        \end{figure}

        Practically, however, the workload at each MCMC step is always shared among a finite number of processors which we denote by $\nproc$. In such case, the total fractional contribution ($C_i$) of the $i$th processor (out of $\nproc$ processors) to the construction of the entire ParaDRAM compact chain is the sum of the probabilities of the occurrences of all acceptances due to the $i$th processor in the simulation,

        \begin{eqnarray}
            C_i
            &\equiv& \pi( \text{acceptance} ~|~ i, \nproc ) \label{eq:processContribution0} \\
            &=& \sum_{j=0}^{+\infty} \geoprob( \text{acceptance} ~|~ k = j\times \nproc + i ) \label{eq:processContribution1} \\
            &=& \maccr ~ \sum_{j=0}^{+\infty} (1-\maccr)^{(j\times \nproc + i-1)} \label{eq:processContribution2} \\
            &=& \maccr ~ (1-\maccr)^{(i-1)} ~ \sum_{j=1}^{+\infty} \big[(1-\maccr)^\nproc\big]^{(j-1)} \label{eq:processContribution3} \\
            &=& \frac{ \maccr ~ (1-\maccr)^{(i-1)} }{ 1 - (1-\maccr)^\nproc } ~ \big( 1 - \big[ (1-\maccr)^\nproc \big]^{j\rightarrow+\infty} \big) \label{eq:processContribution4} \\
            &=& \frac{ \maccr ~ (1-\maccr)^{(i-1)} }{ 1 - (1-\maccr)^\nproc }
            ~~,~~ i = 1, \ldots, \nproc ~, \label{eq:processContribution5}
        \end{eqnarray}

        \noindent where \eqref{eq:processContribution4} and \eqref{eq:processContribution5} are derived from the cumulative distribution function of the Geometric distribution.
        Since the occurrence of an acceptance is checked in order from the first (master) process to the last, the first processor has, on average, always the highest contribution to the construction of the MCMC chain, followed by the rest of the processors in order, as implied by \ref{eq:processContribution5} and illustrated in Figure \ref{fig:procContribution}. This means that the overall scaling behavior of a parallel ParaDRAM simulation solely depends on the contribution ($C_1$) of the first processor to the construction of the MCMC chain. The contribution $C_1$ is in turn determined by the average acceptance rate of the simulation as in \eqref{eq:processContribution5}.

        For example, if the MCMC sampling efficiency is $100\%$, then the entire MCMC output is constructed by the contributions of the first processor. By contrast, the lower the sampling efficiency is, the more evenly the simulation workload will be shared among all processors. Quantitatively, the maximum speedup for a given $\nproc$ number of processors and an average $\maccr$ MCMC sampling efficiency can be written as,

        \begin{equation}
            \label{eq:maxSpeedupCurrent}
            S(\nproc) \approx \frac
            {
                T_s + T_p
            }{
                T_s + C_1(\maccr)\times T_p + (\nproc-1) \times T_o
            }
            ~,
        \end{equation}

        Frequently though, the average acceptance rate ($\alpha$) of an MCMC simulation is a wildly-varying dynamic quantity during the simulation. Therefore, instead of using the estimated average MCMC sampling efficiency from the simulation, we infer an effective MCMC sampling efficiency by fitting the Geometric distribution of \eqref{eq:processContribution5} to the contributions of the individual processors to the output chain. In practice, we find that this effective sampling efficiency is frequently slightly larger than the average MCMC sampling efficiency defined as the ratio of the number of accepted states to the full length of the generated (pseudo)-Markov Chain. Figure \ref{fig:PredictedSpeedupActualSpeedup} compares the simulation speedup predicted in the post-processing section of the ParaDRAM algorithm with the actual simulation speedup, for a range of processor counts.
        \newpar

\section{Example Results}
\label{sec:results}

    A wide range of mathematical test objective functions exist with which the performance of the ParaDRAM algorithm can be benchmarked. The presentation of all examples goes beyond the scope of this manuscript. For illustration purposes, here we present the results for a popular multi-modal example test objective function known as the Himmelblau's function \citep{himmelblau1972applied}. This function is frequently used in testing the performance of optimization algorithms and is defined as,

    \begin{equation}
        \label{eq:Himmelblau}
        f_H(x,y) = (x^2 + y - 11)^2 + (x + y^2 -7)^2 ~,
    \end{equation}

    \noindent with one local maximum at,

    \begin{equation}
        \label{eq:HimmelblauMax}
        f_H(-0.271,-0.923)\approx181.617 ~,
    \end{equation}

    \noindent and four identical local minima located at,

    \begin{equation}
        \label{eq:HimmelblauMin}
        f_H(3.0, 2.0) \approx f_H(-2.805, 3.131) \approx f_H(-3.779, -3.283) \approx f_H(3.584, -1.848) \approx 0.0 ~.
    \end{equation}

    However, just as with any type of MCMC sampler, the ParaDRAM algorithm explores the maxima of objective functions as opposed to the minima. Therefore, we modify the original Himmelblau's function of \eqref{eq:Himmelblau} by adding a small value of $0.1$ to the function and inverting the entire new function, such that all four minima become maxima and remain well-defined the logarithm of the function is computed and passed to the ParaDRAM algorithm. This simulation can be performed in any of the programming language environments that are currently supported by the ParaMonte library. For brevity, here we suffice to presenting only the simulation codes and results generated in  the MATLAB scientific computing language.

    A very simple implementation of this simulation in MATLAB is provided below,

{\small
\begin{lstlisting}[language=octave]
getLogFunc = @(x) -log( (x(1)^2 + x(2) - 11)^2 + (x(1) + x(2)^2 - 7)^2 + 0.1 );
pm = paramonte();
pmpd = pm.ParaDRAM();
pmpd.runSampler(2, getLogFunc);
\end{lstlisting}
}

    The above minimal code defines the natural logarithm of the 2-dimensional Himmelblau's function as a MATLAB anonymous (Lambda) function named \texttt{getLogFunc}, then generates an instance of the \texttt{paramonte} class named \texttt{pm}, from which an instance of the \texttt{ParaDRAM} class is derived and named \texttt{pmpd}. Then the \texttt{runSampler()} method of the \texttt{ParaDRAM} class is called to sample the objective function represented by \texttt{getLogFunc}. All simulation specifications for this sampling problem that are not predefined by the user, will be automatically set to appropriate default values by the ParaDRAM algorithm. Once the simulation is finished, the post-processing tools that are shipped with the ParaMonte-MATLAB library can seamlessly parse, analyze, and visualize the output of the simulation.

    \begin{figure}[t!]
        \centering
        \includegraphics[width=\textwidth]{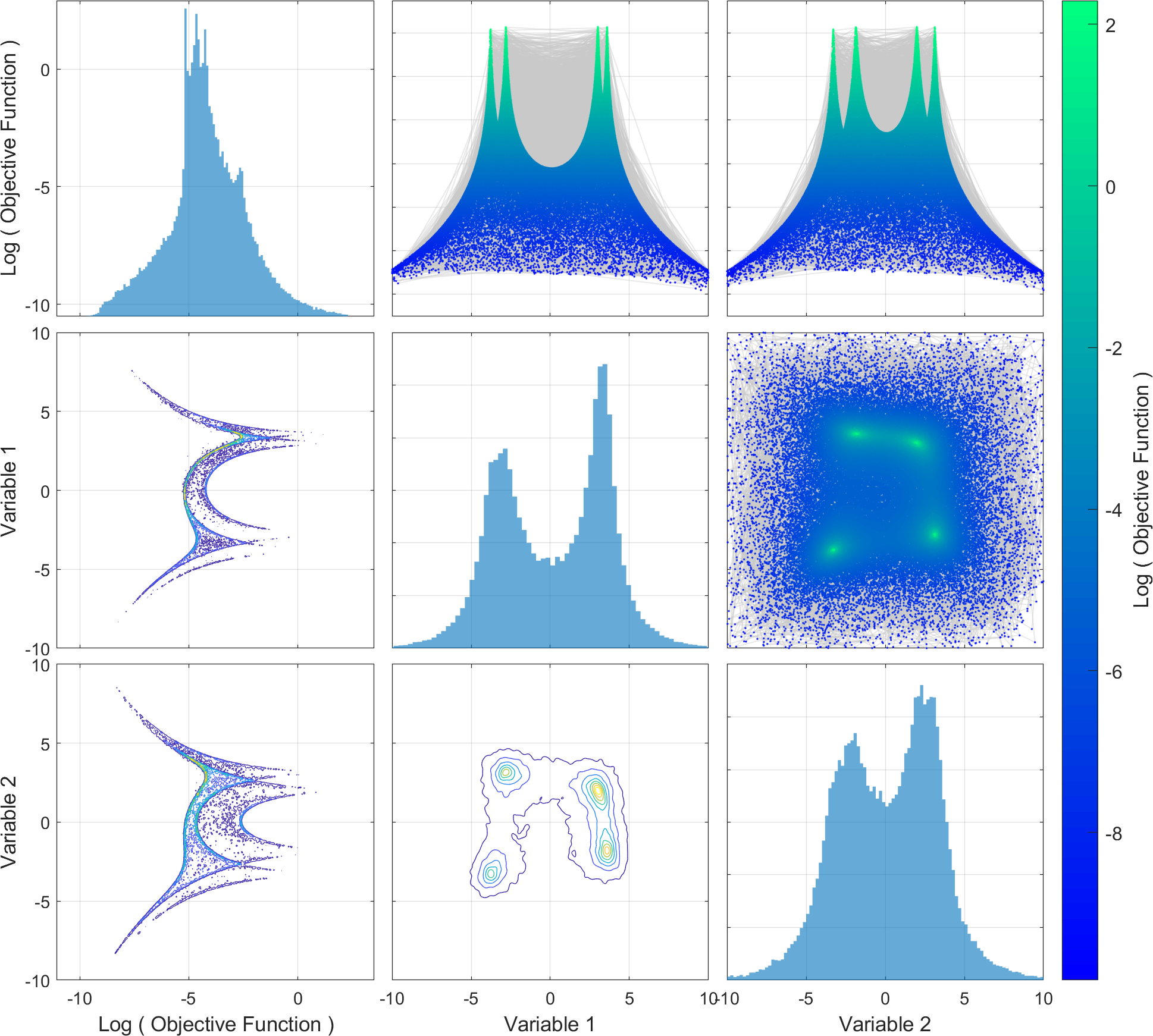}
        \caption{An illustration of the ParaDRAM simulation output for the problem of sampling Himmelblau's function. The figure data consists of pairs of the Himmelblau's function value and its two input variables, plotted against each other. Only the uniquely-visited states in the domain of Himmelblau's function are shown in the plots. The lower triangle of the plot represents the density contour maps of the sampled points, whereas the upper triangle contains line-scatter plots of pairs of variables, color-coded by the natural logarithm of Himmelblau's function. The mono-color gray lines connect the sequence of points in the chain together. The diagonal plots represent the distributions of the uniquely-visited states within the domain of Himmelblau's function. \label{fig:himmelblauGridPlot}}
    \end{figure}

    \begin{figure}[t!]
        \centering
        \begin{subfigure}[t]{0.48\textwidth}
            \centering
            \includegraphics[width=\textwidth]{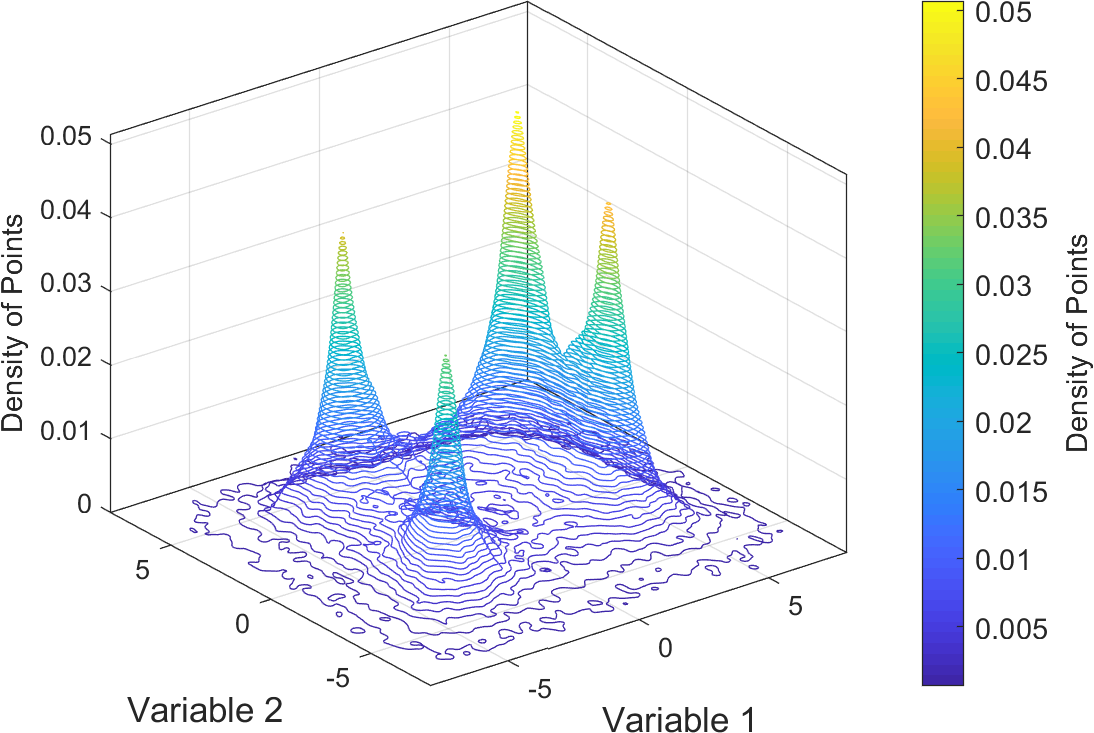}
            \caption{The 3D density contour map.} \label{fig:himmelblauContour3}
        \end{subfigure}
        \hfill
        \begin{subfigure}[t]{0.48\textwidth}
            \centering
            \includegraphics[width=\textwidth]{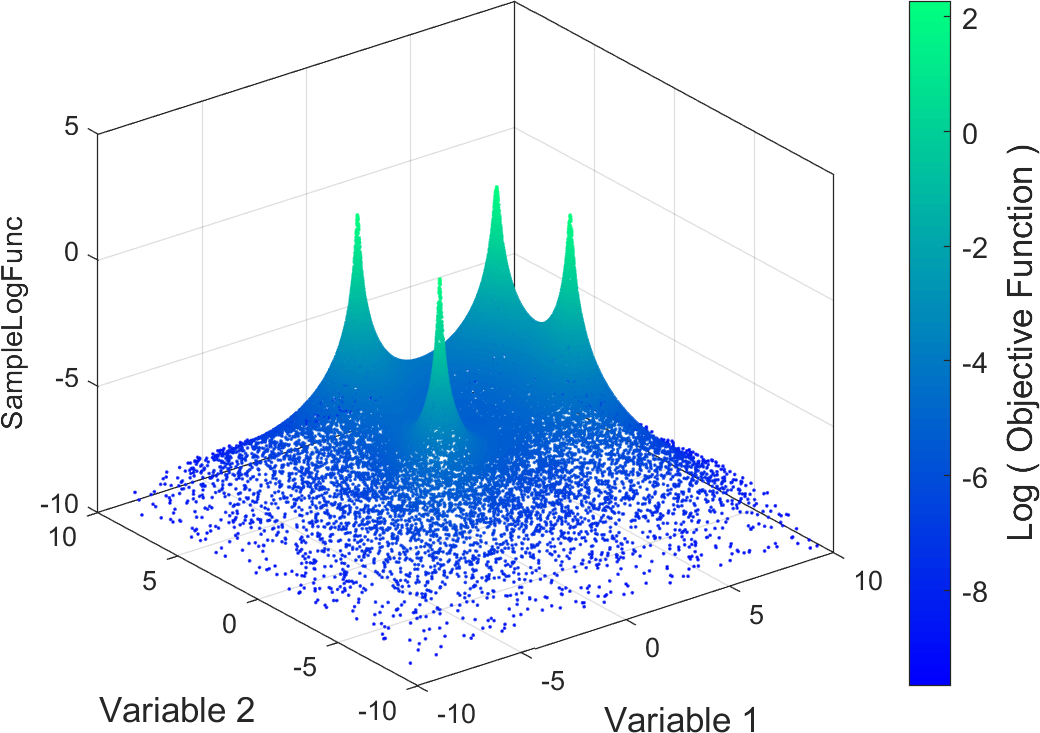}
            \caption{The 3D scatter plot.} \label{fig:himmelblauScatter3}
        \end{subfigure}
        \caption{{\bf(a)} An 3D contour map of the ParaDRAM simulation output for the problem of sampling Himmelblau's function. The figure data consists of the density map of the set of all uniquely-visited states by the ParaDRAM sampler within the domain of the objective function. {\bf(b)} A 3D scatter plot of the set of uniquely-visisted states by the ParaDRAM sampler within the domain of Himmelblau's function. All plots are generated via the visualization tools that automatically ship with ParaMonte-MATLAB library.\label{himmelblau3D}}
    \end{figure}

    Figure \ref{fig:himmelblauGridPlot} illustrates a grid-plot of the uniquely-visited points by the ParaDRAM sampler. A better visualization of the density of the uniquely-visited states within the domain of Himmelblau's function is given in Figure \ref{fig:himmelblauContour3}. The ParaDRAM visualization toolbox uses the linear-diffusion-process kernel density estimation method of \citet{botev2010kernel} to generate the 2D and 3D contour plots. A 3-dimensional visualization of the structure of Himmelblau's function using the uniquely-visited sates by the ParaDRAM algorithm is given in Figure \ref{fig:himmelblauScatter3}.

    \begin{figure}[t!]
        \centering
        \begin{subfigure}[t]{0.48\textwidth}
            \centering
            \includegraphics[width=\textwidth]{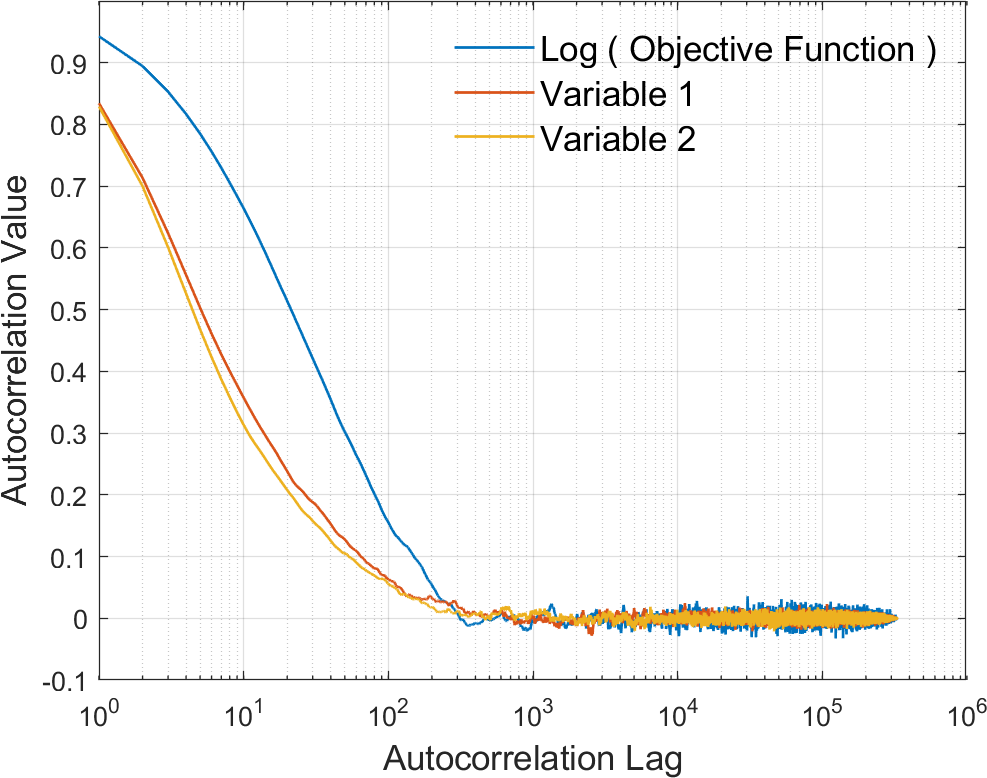}
            \caption{The MCMC autocorrelation.} \label{fig:himmelblauAutoCorrMCMC}
        \end{subfigure}
        \hfill
        \begin{subfigure}[t]{0.48\textwidth}
            \centering
            \includegraphics[width=\textwidth]{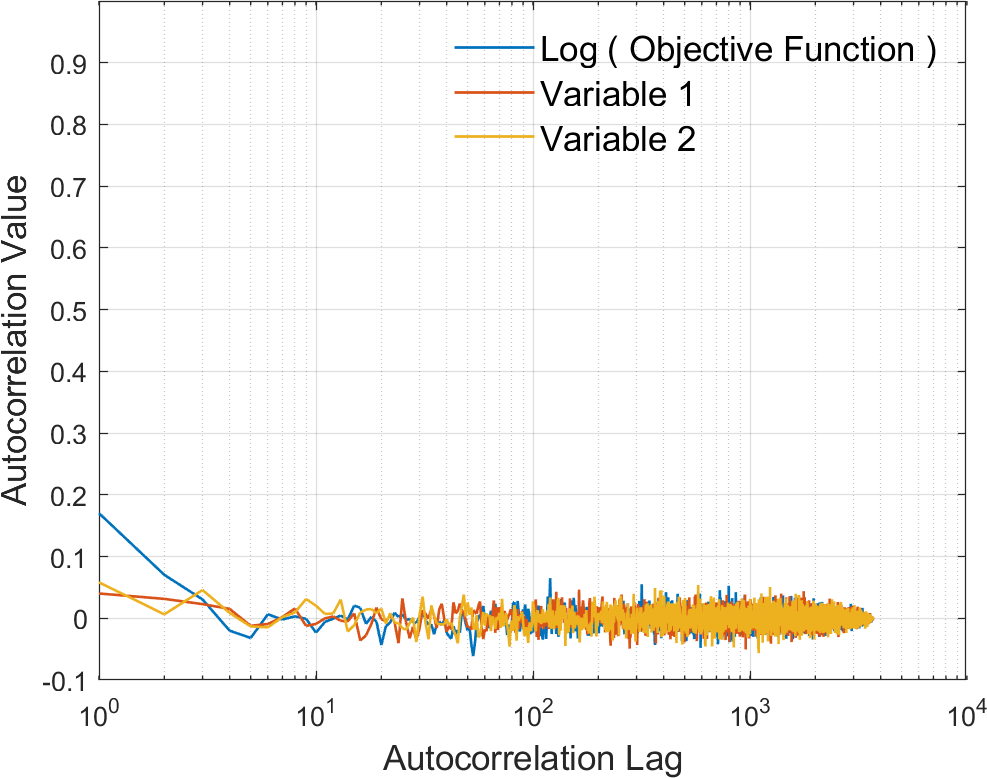}
            \caption{The refined-sample's autocorrelation.} \label{fig:himmelblauAutoCorrSample}
        \end{subfigure}
        \caption{{\bf(a)} An illustration of the autocorrelation in the individual variables of the output MCMC chain in the simulation of Himmelblau's function. {\bf(b)} An illustration of the residual autocorrelation in the individual variables of the final output refined sample in the simulation of Himmelblau's function. By default, the ParaDRAM algorithm performs an aggressive and recursive series of MCMC refinements aimed at removing any traces of autocorrelation in the final refined output sample by the ParaDRAM sampler.\label{fig:himmelblauAutoCorr}}
    \end{figure}

    As mentioned in \S\ref{sec:paradram:refinement}, the ParaDRAM algorithm performs an aggressive series of sample refinements on the generated Markov chains such that no residual autocorrelation remains in the final output sample. Figure \ref{fig:himmelblauAutoCorr} compares the amount of autocorrelation in the original output MCMC chain from the ParaDRAM sampler with the residual autocorrelation in the final refined sample. By default, the ParaDRAM algorithm repeatedly and aggressively refines the output MCMC chain, for as long as needed, such that no traces of autocorrelations remain in the resulting final sample by the algorithm.
    \newpar

    \subsection{Monitoring the dynamic adaptation of the proposal distribution of the ParaDRAM sampler}
    \label{sec:results:benchmark}

        In \S\ref{sec:paradram:diminishing} we argued for necessity of ensuring and monitoring the diminishing adaptation condition of the adaptive algorithms, including the ParaDRAM sampler. Therein, we offered a solution for the dynamic monitoring of the changes in the proposal distribution via an adaptation measure whose value is limited to the range $[0,1]$. Figures \ref{fig:himmelblauProposalDynamics3d} and \ref{fig:himmelblauProposalDynamics2d} display the dynamic evolution of the proposal distribution of the ParaDRAM routine for the problem of sampling Himmelblau's function. The continuous adaption of the proposal is visualized by the changes in the covariance matrix of the proposal distribution throughout the entire simulation. It is evident from the plots that the proposal adaptation eventually diminishes as desired.
        \newpar

        The amount of adaptation in the proposal distribution can be further quantified by the adaptation measure defined in \S\ref{sec:paradram:diminishing}. Figure \ref{fig:himmelblauAdaptationMeasure} illustrates the monotonically-decreasing adaptation of the bivariate Normal proposal distribution used in the sampling of Himmelblau's function via the ParaDRAM sampler. As part of the output chain file, the ParaDRAM algorithm continuously outputs and monitors the diminishing adaptation condition of the DRAM algorithm to ensure the asymptotic ergodicity and the Markovian properties of the resulting output chain. The power-law decay of the adaptation-measure seen in Figure \ref{fig:himmelblauAdaptationMeasure} is exactly the kind of diminishing adaptation condition one would hope to see in ParaDRAM simulations.

    \subsection{Performance benchmarking of the MPI and PGAS parallelism paradigms}
    \label{sec:results:benchmark}

        \begin{figure}[t!]
            \centering
            \begin{subfigure}[t]{0.48\textwidth}
                \centering
                \includegraphics[width=\textwidth]{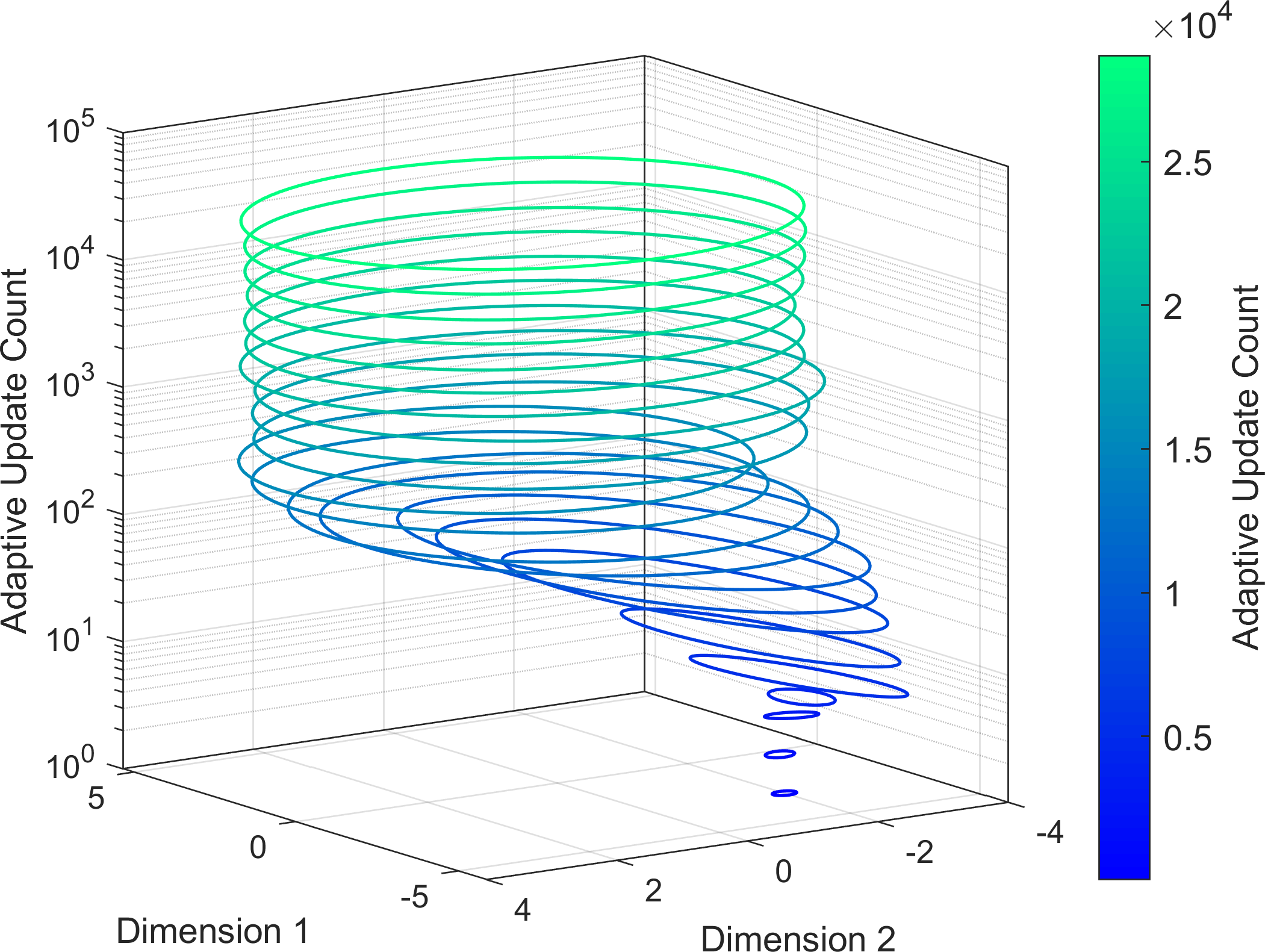}
                \caption{The proposal covariance dynamics in 3D.} \label{fig:himmelblauProposalDynamics3d}
            \end{subfigure}
            \hfill
            \begin{subfigure}[t]{0.48\textwidth}
                \centering
                \includegraphics[width=\textwidth]{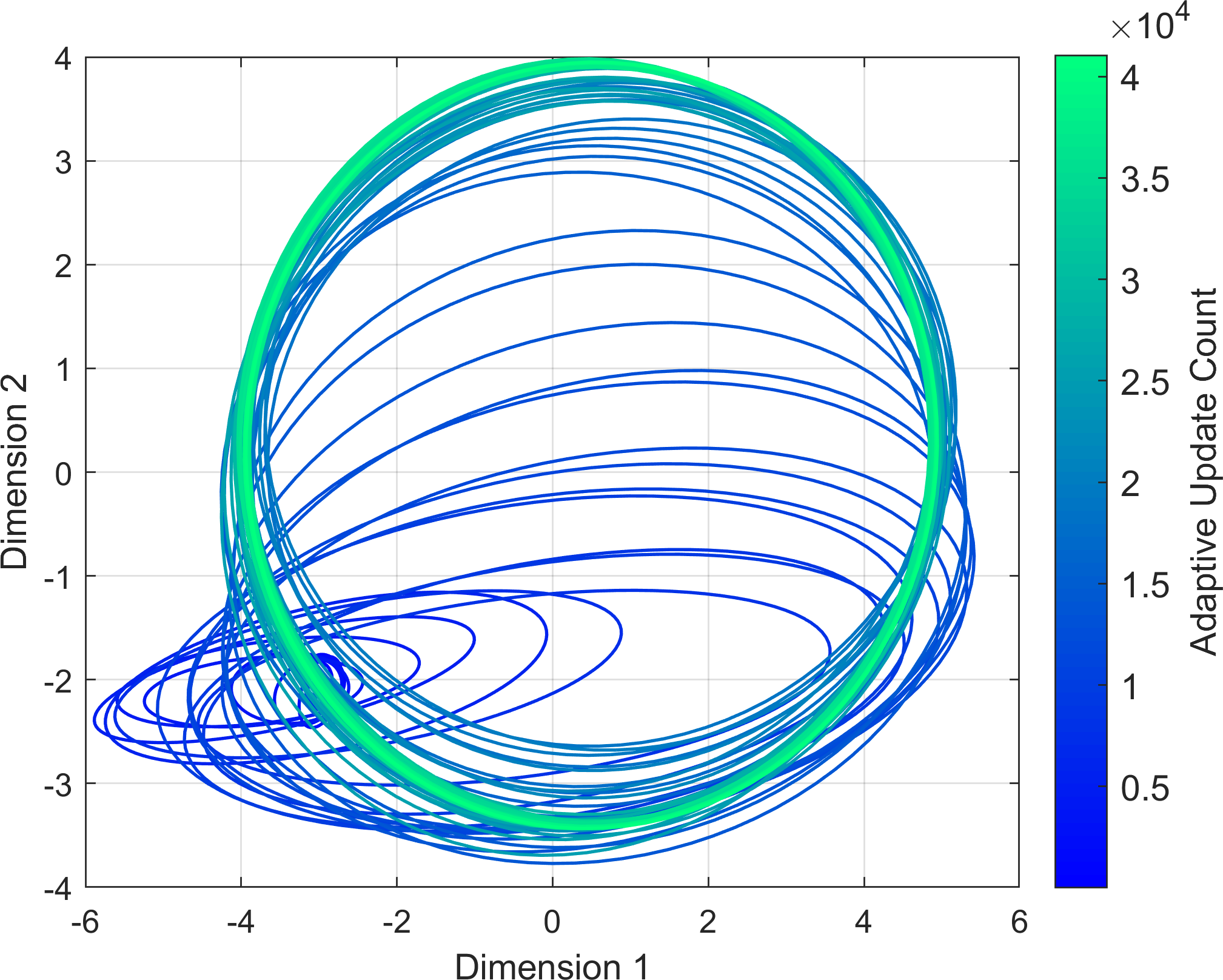}
                \caption{The proposal covariance dynamics in 2D.} \label{fig:himmelblauProposalDynamics2d}
            \end{subfigure}
            \hfill
            \begin{subfigure}[t]{0.48\textwidth}
                \centering
                \includegraphics[width=\textwidth]{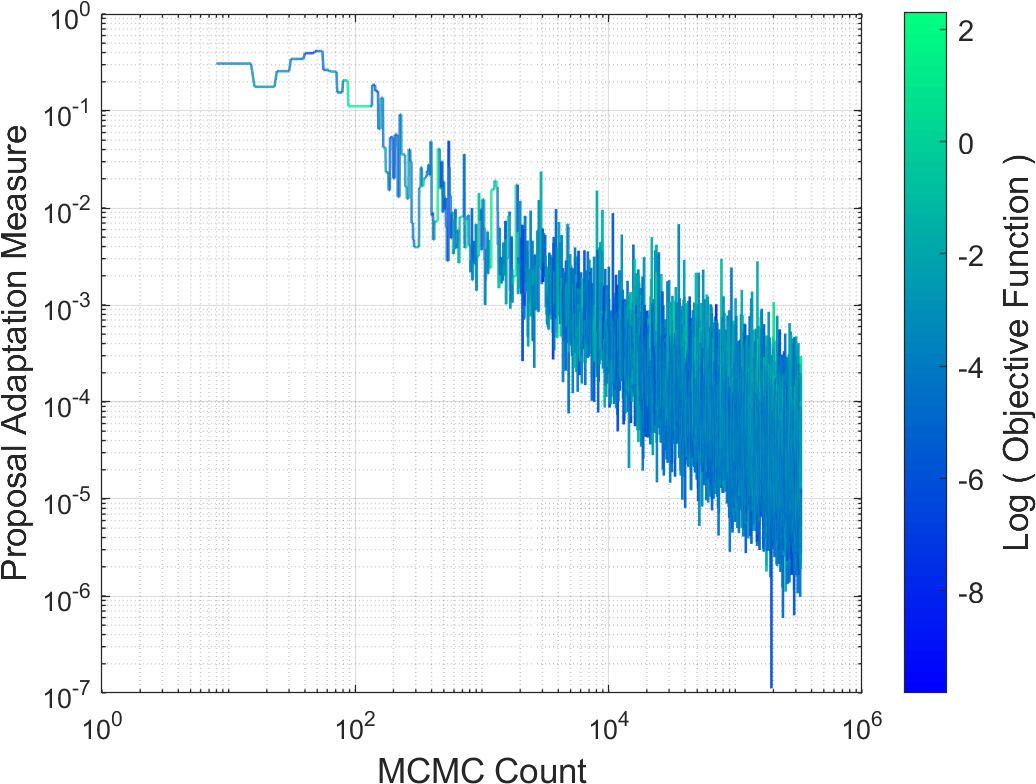}
                \caption{The diminishing adaptation criterion.} \label{fig:himmelblauAdaptationMeasure}
            \end{subfigure}
            \hfill
            \begin{subfigure}[t]{0.48\textwidth}
                \centering
                \includegraphics[width=\textwidth]{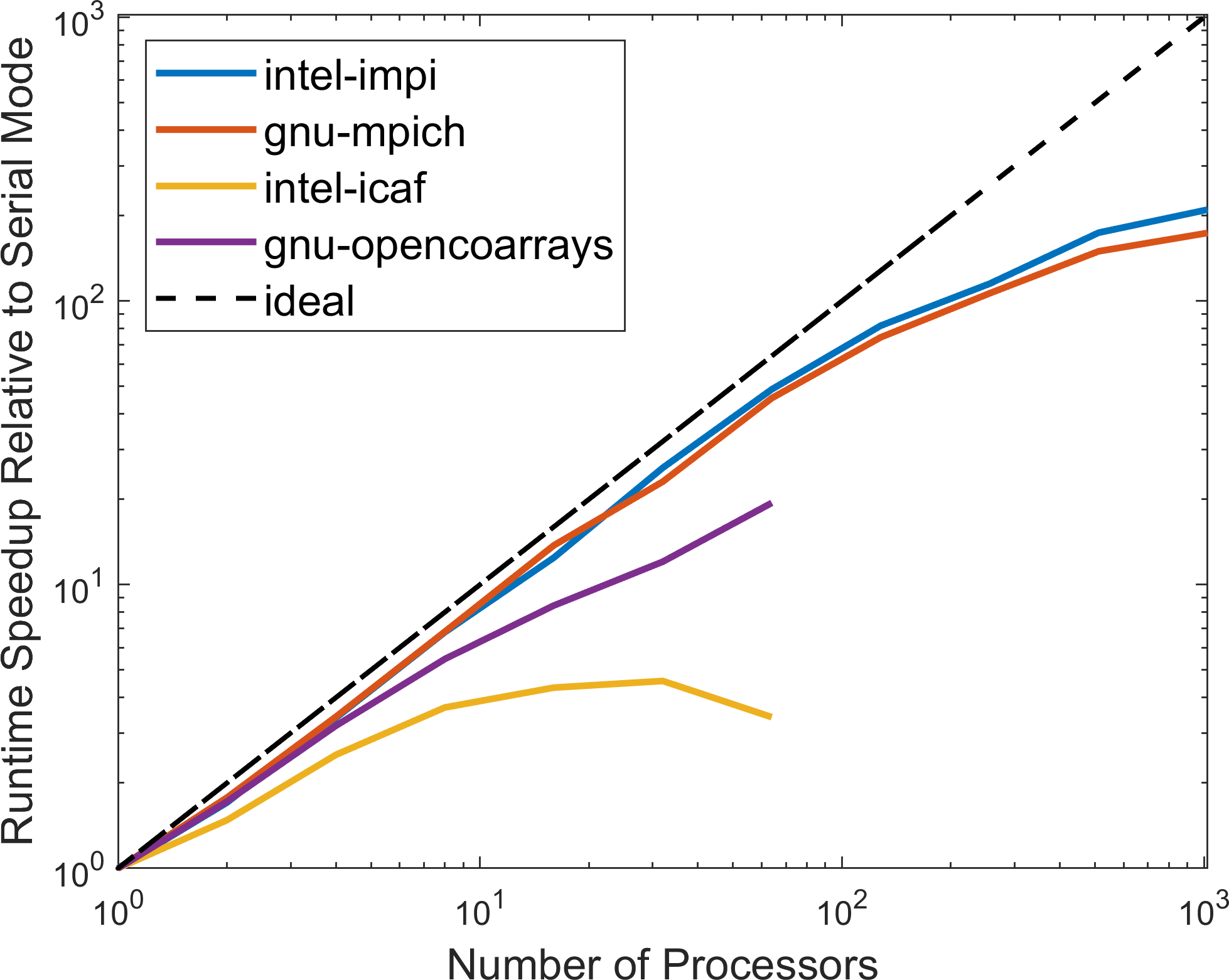}
                \caption{Example strong scaling results: MPI vs. PGAS.} \label{fig:benchmarkParallel}
            \end{subfigure}
            \caption{
                {\bf(a)}
                    An illustration of the 3-dimensional dynamic adaptation of the covariance matrix of the bivariate Normal proposal distribution of the ParaDRAM sampler for the problem of sampling Himmelblaus's function.
                {\bf(b)}
                    An illustration of the 2-dimensional dynamic adaptation of the covariance matrix of the bivariate Normal proposal distribution of the ParaDRAM sampler for the problem of sampling Himmelblaus's function.
                {\bf(c)}
                    An illustration of the diminishing adaptation of the proposal distribution of the ParaDRAM sampler for the problem of sampling Himmelblaus's function. As explained in \S\ref{sec:paradram:diminishing}, the monotonically-decreasing adaptivity observed in the plot ensures the asymptotic ergodicity and Markovian property of the resulting output ParaDRAM chain.
                {\bf(d)}
                    An illustration of the strong-scaling results for parallel ParaDRAM simulations using the two different parallelization paradigms implemented in ParaDRAM: 1. the Message Passing Interface (MPI) via the Intel MPI (impi) and MPICH libraries and, 2. the Partitioned Global Address Space (PGAS) via Intel Coarray Fortran (icaf) and OpenCoarrays library. See \S\ref{sec:discussion} for an explanation of the performance differences between the strong-scaling results of the PGAS- and MPI- parallelized ParaDRAM simulations.}
        \end{figure}

         Given the multiple different parallelism paradigms currently implemented in the ParaDRAM algorithm, it may be of interest to user of the library to know which parallelism paradigm and/or perhaps what compilers or parallism library implementations yield the best simulation performances. Figure \ref{fig:benchmarkParallel}, illustrates the performance benchmarking of the ParaDRAM algorithm for an example 4-dimensional multivariate Normal target density function. Given the simplicity of such sampling problem, the time-cost of calling this objective function was artificially increased so that a more accurate and clear comparison could be made between the strong-scaling results for the MPI and PGAS parallelism paradigms.
         \newpar

         We obtained and compared the results for the MPI and PGAS parallelism methods using compilers from two different vendors: the Intel and the GNU compiler suites. In the case of the MPI parallelism, the Intel MPI and the MPICH MPI libraries were used respectively. In the case of the PGAS parallelism, the Intel Coarray and the OpenCoarrays \citep{numrich1998co, fanfarillo2014opencoarrays} libraries were used respectively. Based on the benchmarking results presented in Figure \ref{fig:benchmarkParallel}, the PGAS parallelism as implemented in ParaDRAM performs inferior to the MPI implementation. There are potentially two reasons for such performance difference between the two parallelism paradigms in ParaDRAM,

         \begin{enumerate}
            \item
                The current version of the ParaDRAM algorithm does not fully exploit the unique readily-available RMA-communication features of the Coarray libraries. This is partly due to the lack of support for the advanced RMA features in the Coarray libraries when the ParaMonte library was originally developed. Recently, however, many new advanced RMA communication features of the Coarray parallelism paradigm have been implemented by multiple open-source and commercial compilers, including the GNU, Intel, and NAG compiler suites. Therefore, we anticipate that a future re-implementation of the PGAS Coarray parallelism in ParaDRAM via the newly-available RMA features will resolve some of the discrepancies observed between the performances of the MPI and PGAS parallelization of the ParaDRAM sampler.
            \item
                The currently-available MPI libraries are highly optimized and mature, while the Coarray PGAS libraries have only recently become available.
         \end{enumerate}

\section{Discussion}
\label{sec:discussion}

    Over the past 3 decades, the popularity and the utilities of Monte Carlo simulations has grown exponentially in a wide range of scientific disciplines. In particular, the Markov Chain Monte Carlo (MCMC) techniques have become indispensable tools for predictive computing and uncertainty quantification. In this work, we presented the ParaDRAM algorithm, a high-performance implementation of the Delayed-Rejection Adaptive Metropolis-Hastings (DRAM) algorithm of \citet{haario2006dram}. The DRAM algorithm is one of the most popular and most successful adaptive MCMC techniques available in the literature that has proven to dramatically outperform the traditional MCMC sampling techniques.
    \newpar

    The presented ParaDRAM algorithm is part of the ParaMonte open-source Monte Carlo simulation library, available at \url{https://github.com/cdslaborg/paramonte}. The library is currently comprised of approximately 130,000 lines of codes primarily in written the C, Fortran, MATLAB, Python, as well as the Bash, Batch, Cmake scripting and build languages. The majors goals in the development of the ParadRAM algorithm have been to bring simplicity, full-automation, comprehensive reporting, and automatic fully-deterministic restart functionality to the inherently stochastic Monte Carlo simulations. In addition, we have careful to design a unified Application Programming Interface (API) to a wide range of popular programming languages in the scientific community, such that the syntax of calling and setting up the ParaDRAM sampler remains almost the same across all programming language environments. Notably, we aimed to achieve the aforementioned goals without compromising the high-performance and the parallel scalability of the algorithm.
    \newpar

    To ensure the scalability of parallel ParaDRAM simulations, from personal laptops to the world-class supercomputers, we have adopted and implemented the MPI and PGAS distributed-memory parallelism paradigms in this library. Remarkably, we have been careful to not require any parallel-coding effort or experience from the user in order to build and run parallel ParaDRAM simulations, from any programming language environment.
    \newpar

    To maintain the high-performance of the library, we also described in this manuscript an efficient compact storage method for the output MCMC chains from the ParaDRAM simulations. This approach as detailed in \S\ref{sec:api:storage} enables us to maximize the library's IO performance and minimize the external and internal memory-footprints of the library, without any compromise in the automatic fully-deterministic restart functionality feature of the ParaDRAM algorithm in parallel or in serial simulations. We also discussed in \S\ref{sec:paradram:diminishing}, a novel technique to automatically and dynamically monitor and ensure the diminishing adaptation criterion of the DRAM algorithm.
    \newpar

    The ParaDRAM sampler library is currently being actively developed and expanded with new sampling capabilities. Further planned enhancements include but are not limited to: 1. increasing the accessibility of the ParaDRAM library from other popular programming languages. There are currently ongoing efforts to include support for the Java, Julia, Mathematica, and R programming languages. 2. minimizing the effects of improper processor load-balance on the overall performance of the parallel simulations. The recent enhancements and additions to the RMA communication facilities within Coarray-PGAS parallelism paradigms will be a great aid toward achieving this goal.
    \\\\

{\bf Acknowledgements} \\

We thank the Texas Advanced Computing Center (TACC) for providing the parallel computing resources for the development and testing of the ParaMonte/ParaDRAM library presented in this manuscript. 

\normalsize
\bibliographystyle{spbasic} %apa}
\bibliography{../../../libtex/all}

\end{document}